\documentstyle[11pt,epsf]{article}
\def\ft#1#2{{\textstyle{{\scriptstyle #1}\over {\scriptstyle #2}}}}
\def\fft#1#2{{#1 \over #2}}
\def\sst#1{{\scriptscriptstyle #1}}

\def\del{\partial}
\def\a{\alpha}
\def\im{{\rm i}}
\def\oneone{\rlap 1\mkern4mu{\rm l}}

\def\dt{{\tilde d}}
\def\beq{\begin{equation}}
\def\eeq{\end{equation}}
\def\bea{\begin{eqnarray}}
\def\eea{\end{eqnarray}}

\def\dalemb#1#2{{\vbox{\hrule height .#2pt
        \hbox{\vrule width.#2pt height#1pt \kern#1pt
                \vrule width.#2pt}
        \hrule height.#2pt}}}
\def\square{\mathord{\dalemb{6.8}{7}\hbox{\hskip1pt}}}

\begin{document}
\topmargin 0pt
\oddsidemargin 5mm
\begin{titlepage}
\begin{flushright}
CTP TAMU-31/95\\
Imperial/TP/94-95/56\\
SISSA 97/95/EP\\
hep-th/9508042\\
\end{flushright}
\vspace{1.5truecm}
\begin{center}
{\bf {\Large Stainless Super $p$-branes}}

\vspace{1.5truecm}

{\large H. L\"u$^{\sst{\clubsuit\heartsuit}}$\footnote{Supported in part by the
U.S. Department of Energy, under grant DE--FG05--91--ER40633.},
C.N. Pope$^{\sst{\clubsuit\heartsuit} 1,}$\footnote{Supported in part by the
EC Human Capital and Mobility Programme under contract ERBCHBGCT920176.},
E. Sezgin$^{\sst{\clubsuit\spadesuit}}$\footnote{Supported in
part by the National
Science Foundation, under grant PHY--9411543.}
and K.S. Stelle$^{\sst{\diamondsuit\heartsuit}}$\footnote{Supported in
part by the
Commission of the European Communities under contract SC1*--CT92--0789.}}

\vspace{1.1truecm}

{\small {\Large $\sst{\clubsuit}$}  Center for Theoretical Physics,
Texas A\&M University,
                College Station, TX 77843-4242, USA}

{\small {\Large $\sst{\diamondsuit}$} The Blackett Laboratory,
Imperial College, Prince Consort
Road, London SW7 2BZ, UK}

{\small {\Large $\sst{\heartsuit}$} SISSA, Via Beirut No. 2-4, 34013
Trieste, Italy}

{\small {\Large $\sst{\spadesuit}$} ICTP, Strada Costiera 11, 34013
Trieste, Italy}

\end{center}

\vspace{1.0truecm}

\begin{abstract}
\vspace{1.0truecm}
The elementary and solitonic supersymmetric $p$-brane solutions to
supergravity theories form families related by dimensional reduction,
each headed by a maximal (`stainless') member that cannot be isotropically
dimensionally oxidized into higher dimensions. We find several new families,
headed by stainless solutions in various dimensions $D\le 9$. In some
cases, these occur with dimensions $(D,p)$ that coincide with those of
descendants of known families, but since the new solutions are stainless,
they are necessarily distinct. The new stainless supersymmetric solutions
include a 6-brane and a 5-brane in $D=9$, a string in $D=5$, and particles
in all dimensions $5\le D\le 9$.
\end{abstract}
\end{titlepage}
\newpage
\pagestyle{plain}
\setcounter{footnote}{0}
\section{Introduction}

     Since the discovery of $p$-brane theories with manifest spacetime
supersymmetry \cite{pol,berg}, it has become increasingly clear that there
is a close relationship between such theories and the set of soliton-like
solutions to supergravity theories \cite{town}.  All the known supersymmetric
$p$-brane theories achieve a matching of the on-shell world-volume bosonic
and fermionic degrees of freedom, by virtue of a local
fermionic symmetry known as $\kappa$ symmetry.  This symmetry compensates
for the excess of fermionic over bosonic degrees of freedom by gauging away
half of the former.  The consistency of $\kappa$ symmetry with spacetime
supersymmetry places severe constraints on the spacetime dimension $D$ and
the world volume dimension $d=p+1$ \cite{achu}.  Four classic families of
super $p$-branes were found to satisfy the consistency criterion.  The
members within each family are related by a process of double dimensional
reduction \cite{dhis}, in which both the spacetime and the world volume are
simultaneously compactified on a circle, and the dependence on the extra
direction is dropped in each space.  Thus the classic super $p$-branes may
be classified by giving the maximal-dimensional member of each of the four
families.   These occur in $(D,d)= (11,3)$, $(10,6)$, $(6,4)$ and $(4,3)$.
On a plot or `brane scan' of $D$ {\it vs} $d$, the additional $p$-branes
obtained by double dimensional reduction lie on the North-east/South-west
diagonal lines descending from the maximal cases.

    The idea that a super $p$-brane could be viewed as a long-wavelength
description of a topological defect in a supersymmetric theory originated
in the construction of the supermembrane in $D=4$ \cite{pol}.  This
supermembrane occurs as a kink solution of a $D=4$ chiral
scalar supermultiplet theory with a potential giving a degenerate vacuum.  A
crucial feature of this solution is that half the original supersymmetry
is left unbroken.  This partial breaking of supersymmetry is also a general
feature of all the subsequently-discovered $p$-brane solitons.

     Another feature of super $p$-branes became clear with the
curved-superspace construction of the $D=11$ supermembrane action in
\cite{berg}, and its generalisations to the other classic super $p$-branes.
This new feature was the occurrence of integrability conditions on the
supergravity background that are required for the existence of the
world-volume $\kappa$ symmetry.  In the case of the $D=11$ supermembrane,
and of the type IIA string, related to it by double dimensional reduction,
these integrability conditions imply the full set of supergravity field
equations \cite{berg,dhis}.

     The association of super $p$-branes to supergravity is also natural
because the supersymmetric $p$-branes can be viewed as the natural `matter'
sources for the corresponding supergravity theories.  A very specific
r\^ole in this association is played by the antisymmetric tensor field
strengths, whose gauge potentials couple directly to the
$(p+1)$-dimensional world volumes.  In the coupled solutions of super
$p$-branes and their corresponding supergravity backgrounds, the
backgrounds are naturally singular on the $p$-brane world volumes, which
can act like delta-function sources.  These singularities may or may not be
clothed by horizons, depending upon the circumstances.  Such singular
supergravity solutions are called `elementary,' in distinction to
the non-singular `solitonic' solutions described previously.

     The association of $p$-branes with singular supergravity solutions was
made concrete with the explicit construction of superstring solutions in the
case of $N=1$, $D=10$ supergravity \cite{dabl}.  These solutions preserve
half of the original supersymmetry, and consequently they saturate a
Bogomol'ny bound on the energy density.  Subsequently, an analogous
elementary membrane solution of $D=11$ supergravity was found \cite{dust}.
Many further solutions of supergravity theories have also been found, both
for elementary $p$-branes \cite{pew1} and for solitonic $p$-branes
\cite{pew2}.  (There are also solitonic solutions in supergravity theories
coupled to Yang-Mills, such as that based upon Yang-Mills instantons, and
corresponding to the heterotic string \cite{pew3}.)

     The multiplicity of elementary and solitonic $p$-brane solutions to
supergravity theories, covering many more values of $(D,d)$ than the
classic $\kappa$-symmetric points on the brane scan, suggests that the
original classification needs to be generalised.  Leaving aside for the
moment the problem of formulating more general $\kappa$-symmetric actions,
it is worthwhile to try to find the general pattern of elementary and
solitonic $p$-brane solutions in supergravity theories.

     Many supergravity theories in $D\le10$ dimensions can be obtained
from $D=11$ supergravity by Kaluza-Klein dimensional reduction, in which a
consistent truncation of the higher-dimensional to the lower-dimensional
theory is made.  Since the truncation is consistent, it follows that
solutions of the lower-dimensional theory are also solutions of the
higher-dimensional one.  This lifting of solutions to the higher dimension
is known as dimensional oxidation.  In some cases, an elementary or
solitonic brane solution in the lower dimension oxidizes to another
elementary or solitonic brane solution in the higher dimension.   The
ability to view an oxidized brane solution as itself being a brane solution
depends upon whether the isotropicity of the lower-dimensional solution
extends to an isotropicity in the higher-dimensional sense.  For the
isotropicity to extend, the extra coordinate of the higher-dimensional
spacetime must either become isotropically grouped with the $p$-brane
coordinates of the lower dimension, making a $(p+1)$-brane, or else it must
become isotropically grouped with the coordinates of the transverse space,
making a $p$-brane in the higher dimension.  As we shall show later, the
latter can never happen within the framework of Kaluza-Klein dimensional
reduction.  The former, on the other hand, can occur under certain
circumstances.  This is the direct analogue, at the level of solutions to
supergravity theories,  of the process of double dimensional reduction of
$p$-brane actions \cite{dhis}.  Just as for those actions, it is useful in
classifying the brane solutions to distinguish between the ones that can be
oxidized to isotropic brane solutions of a higher-dimensional supergravity
theory, and those that cannot be isotropically oxidized.  We shall call the
former solutions `rusty,' and the latter solutions `stainless.'  Thus when
constructing a brane scan of supergravity solutions, one may omit the rusty
solutions, which are simply the Kaluza-Klein descendants of stainless
solutions in some higher dimension.  The full solution set is thus
characterised by the stainless solutions.

     A frequently-encountered contention in the recent literature is that
the only fundamental brane solutions occur in $D=11$ and $D=10$
supergravities, and that all the others are simply obtained by
dimensional reduction.  In this paper, we shall show that this is not the case,
given our requirement of isotropicity in the oxidation process.\footnote{
An opposite viewpoint is to regard all oxidations of brane solutions as branes
in the higher dimension.  We prefer not to adopt this viewpoint since, if the
isotropicity requirement on the world volume is dropped, the solutions are not
ordinary extended objects, and moreover it would not then be clear what degree
of anisotropicity should be regarded as acceptable.} In particular, we shall
find new stainless brane solutions to supergravity theories in all $5\le
D\le9$.  (We shall not be concerned in the present paper with supersymmetric
$p$-brane solutions to super Yang-Mills or other rigid supersymmetric
theories.)  Amongst other stainless examples, we shall find a 6-brane and a
5-brane in $D=9$, and a string in $D=5$, none of which are obtainable from
$D=11$ or $D=10$ $p$-brane solutions by dimensional reduction.

\section{Solutions and Kaluza-Klein dimensional reduction}
\subsection{$p$-brane solutions}

      We are concerned with elementary and solitonic solutions
of supergravity theories that admit interpretations as
$p$-branes embedded in spacetime.  These solutions will in general involve the
metric tensor $g_{\sst{MN}}$, a dilaton $\phi$ and an $n$-index
antisymmetric tensor $F_{\sst{M_1\cdots M_n}}$ in $D$ dimensions.  The
Lagrangian for these fields takes the form
\beq
{\cal L} = e R - \ft12 e(\del\phi)^2 - {1\over 2\, n!} e e^{-a \phi} F^2\ ,
\label{boslag}
\eeq
where $e=\sqrt{-g}$ is the determinant of the vielbein.  The equations of
motion are
\bea
\square \phi &=& -{a\over 2 n!}\, e^{-a \phi} F^2\ ,\nonumber\\
R_{\sst{MN}} &=& \ft12 \del_{\sst{M}}\phi\,\del_{\sst{N}}\phi +
S_{\sst{MN}} \ ,\label{eqmo1}\\
\del_{\sst{M_1}} (e e^{-a \phi} F^{\sst{M_1\cdots M_n}}) &=& 0\ ,\nonumber
\eea
where $S_{\sst{MN}}$ is a symmetric tensor given by
\beq
S_{\sst{MN}} = {1\over 2 (n-1)!}\, e^{-a \phi}\,\Big( F^2_{\sst{MN}} -
\fft{n-1}{n(D-2)} F^2 g_{\sst{MN}}\Big) \ .\label{smndef}
\eeq

      The ansatz for the metric for the $D$ dimensional spacetime is given
by \cite{dabl,dkl}
\beq
ds^2 = e^{2A}\, dx^\mu dx^\nu \eta_{\mu\nu} +
       e^{2B}\, dy^m dy^n \delta_{mn}\ ,\label{metrform}
\eeq
where $x^{\mu}$ $(\mu = 0, \ldots, d-1)$ are the coordinates of the
$(d-1)$-brane world volume, and $y^m$ are the coordinates of the
$(D-d)$-dimensional transverse space.   The functions $A$ and $B$ depend
only on $r=\sqrt{y^my^m}$.  Note that the form of the metric ansatz is
preserved under the replacement $r\longrightarrow 1/r$.   The Ricci tensor
for the metric (\ref{metrform}) is given by
\bea
R_{\mu\nu} &=& -\eta_{\mu\nu} e^{2(A-B)} \Big ( A'' + d {A'}^2
+ \dt A' B' + \fft{\dt+1}{r}\, A'\Big)\ ,\nonumber\\
R_{mn} &=& -\delta_{mn} \Big(B'' + d A' B' + \dt {B'}^2 + \fft{2\dt +1}r\, B'
+ \fft{d}r\, A'\Big) \label{ricci}\\
&& -{y^my^n \over r^2} \Big(\dt B'' + d A'' - 2d A'B' + d {A'}^2 -
\dt {B'}^2 -\fft{\dt}{r}\, B' - \fft{d}{r} \, A'\Big)\ ,\nonumber
\eea
where $\dt= D -d - 2$ and a prime denotes a derivative with respect to
$r$.   A convenient choice of vielbein basis for the metric
(\ref{metrform}) is $e^{\underline \mu} = e^{A} d x^\mu$ and $
e^{\underline m} = e^{B} dy^m$, where underlined indices denote tangent
space components.  The corresponding spin connection is
\bea
\omega^{{\underline\mu}{\underline n}} &=& e^{-B}\del_n A\, e^{\underline \mu}
\ ,\qquad \omega^{{\underline \mu}{\underline \nu}} = 0\ ,\nonumber\\
\omega^{{\underline m}{\underline n}} &=& e^{-B} \del_n B\, e^{\underline m}
-e^{-B} \del_m B\, e^{\underline n}\ .\label{spincon0}
\eea

     For the elementary $p$-brane solutions, the ansatz for the
antisymmetric tensor is given in terms of its potential, and takes the form
\cite{dabl}
\beq
A_{\mu_1\ldots\mu_{n-1}} = \epsilon_{\mu_1\ldots\mu_{n-1}} e^C
\ ,\label{eleans}
\eeq
and hence
\beq
F_{m\mu_1\ldots\mu_{n-1}} = \epsilon_{\mu_1\ldots\mu_{n-1}} \del_m e^C\ ,
\label{eleans2}
\eeq
where $C$ is a function of $r$ only.  Here and throughout this paper
$\epsilon_{\sst{M\cdots N}}$ and $\epsilon^{\sst{M\cdots N}}$ are taken to
be the tensor densities of weights $-1$ and 1 respectively, with purely
numerical components $\pm 1$ or $0$.  Note in
particular that they are not related just by raising and lowering indices using
the metric tensor.   The dimension of the world volume is given by $d=n-1$
for the elementary $p$-brane solutions.

     For the solitonic $(d-1)$-brane solutions, the ansatz for the
antisymmetric tensor is \cite{pew2}
\beq
F_{m_1\cdots m_n} = \lambda \epsilon_{m_1\cdots m_n p }\, {y^p\over r^{n+1}}
\ ,\label{solans}
\eeq
where $\lambda$ is a constant.  The power of $r$ is governed by the
requirement that $F$ should satisfy the Bianchi identity.  The dimension of
the world volume is given by $d=D-n-1$ for the solitonic $p$-brane
solutions.

       For both types of solution, the symmetric tensor $S_{\sst{MN}}$
takes the form
\bea
S_{\mu\nu} &=& -\fft{\dt}{2(D-2)}\, S^2\, e^{2(A-B)}
\eta_{\mu\nu} \ ,\nonumber\\
S_{mn} &=& \fft{d}{2(D-2)}\, S^2 \,\delta_{mn} -
\ft12\, S^2\, {y^my^n \over r^2}
\ .\label{smnform}
\eea
The function $S$ is given in the two cases by
\bea
{\rm elementary:}&& S = e^{-\ft12 a\phi - d A + C}\, C'\qquad \ \ \ d=n-1\ ,
\nonumber\\
{\rm solitonic:}&& S = \lambda e^{-\ft12 a\phi - \dt B}\,  r^{-\dt-1}
\qquad d= D-n-1\ .\label{sforms}
\eea

     With these ans\"atze, the equations of motion for the dilaton and the
metric tensor in (\ref{eqmo1}) become
\bea
\phi'' + d A' \phi' + \dt B' \phi' + \fft{\dt +1}r\, \phi' &=&\ft12
\epsilon a S^2\ ,\nonumber\\
A'' + d {A'}^2 + \dt A' B'  + \fft{\dt+1}r \, A' &=& \fft{\dt}{2(D-2)} S^2\
,\nonumber\\
B'' + d A' B' + \dt {B'}^2 + \fft{2\dt +1}r \, B' + \fft{d}r\, A' &=&
-\fft{d}{2(D-2)} S^2 \ ,\label{eqmo2}\\
\dt B'' + d A'' - 2 d A' B' + d {A'}^2 - \dt {B'}^2 -
\fft{\dt}r \, B' - \fft{d}r \, A' + \ft12 {\phi'}^2 &=& \ft12 S^2\ ,\nonumber
\eea
where $\epsilon = 1$ for the elementary ansatz and $\epsilon = -1$
for the solitonic ansatz.  The equation of motion for the field strength
$F$ in (\ref{eqmo1}) is automatically satisfied by the solitonic ansatz
(\ref{solans}), whilst for the elementary ansatz (\ref{eleans}) it gives
rise to the equation
\beq
C'' + C'(C' + \dt B' - dA' - a \phi') + \fft{\dt+1}r \, C' = 0\ .
\label{ceq}
\eeq

    Solutions to the equations of motion (\ref{eqmo2}) and (\ref{ceq}) can
be obtained by making the following ansatz:
\beq
A' =\fft{\epsilon}{\Lambda}\, S\ ,\qquad \phi' = \fft{\epsilon(D-2)a}{\dt}
\,  A'\ ,\label{soluans}
\eeq
where $\Lambda$ is a constant.  By choosing $\Lambda$ such that
\beq
\Lambda^2 = \fft{(D-2)^2 a^2}{\dt^2} + \fft{2d(D-2)}{\dt}\ ,\label{betaeq}
\eeq
one can eliminate the non-linear terms ${A'}^2$, ${B'}^2$ and $A' B'$ from
a linear combination of the last three equations in
(\ref{eqmo2}).\footnote{There are more general solutions of the equations
(\ref{eqmo2}) than those that follow from the ansatz
(\ref{soluans},\ref{betaeq}).  However, as we shall see later, when one
considers supergravity theories the equations implied by requiring that
half the superymmetry be preserved are equivalent to
(\ref{soluans},\ref{betaeq}).}    Then it
is a simple matter to solve the equations; the solution is given by
\bea
B&=&-\fft{d}{\dt} \, A \ , \qquad \phi = \fft{a(D-2)}{\epsilon \dt} \, A
\ ,\nonumber\\
e^{-c A} &=& 1 + \fft{k}{r^{\dt}}\ ,\label{solution1}
\eea
where $k = \epsilon \Lambda\lambda/(2(D-2))$ and $ c= d + a^2
(D-2)/(2\dt)$.  In the elementary case, the function $C$ satisfies the
equation
\beq
(e^C)' = \lambda\,  e^{2cA}\, r^{-\dt -1}\ .\label{csol}
\eeq
In presenting these solutions we have chosen simple values for some
integration constants where no loss of generality is involved.
The solutions (\ref{solution1}) are valid when $d\dt>0$.  For the cases
$d=0$ or $\dt=0$, the solutions can also be straightforwardly obtained; an
example will be given in section 4.2.  Note that the forms of the metrics for
both elementary and solitonic $(d-1)$-branes are the same, although, as we
saw earlier, the solutions are obtained from a $(d+1)$-form antisymmetric
tensor field strength in the former case, and from a $(D-d-1)$-form
antisymmetric tensor field strength in the latter case.

      So far, we have obtained solutions for the bosonic theory described
by the Lagrangian (\ref{boslag}) for arbitrary values of the constant $a$,
and with an antisymmetric tensor of arbitrary degree.   In supergravity
theories, however, there occur antisymmetric tensors of certain specific
degrees only, each with its corresponding specific value of the constant
$a$.   We may summarise the $a$ values arising in supergravity theories as
follows. Without loss of generality, we may discuss all theories in
versions where all antisymmetric tensor field strengths have
degrees $n\le D/2$. The $a$ values are given by
\beq
    a^2 = \Delta -\fft{2d\dt}{D-2}\ ,\label{avalue}
\eeq
where
\beq
d\dt=(n-1)(D-n-1)\ .\label{ndep}
\eeq
Some examples of values of $\Delta$ that arise in supergravity theories
are $\Delta =4$ for $n \ne 2$, and $\Delta = 4$ and 2 for $n=2$. (See
\cite{SS}, where a large class of supergravity theories in various dimensions
can be found.) We shall
discuss the set of $\Delta$ values in more detail in section 4.1. Note that
in cases where there is no dilaton, the solution for the $A$ and $B$
functions that appear in the metric ansatz is precisely given by
(\ref{solution1}) with the value of $a$ taken to be zero.   In this sense
we can assign the value $\Delta = 2d\dt/(D-2)$, which, by eqn
(\ref{avalue}), sets $a=0$,  in a supergravity theory where there is no
dilaton.  For example $\Delta=4$ for the 4-form field strength in $D=11$
supergravity, $\Delta=2$ for the 3-form field strength in $D=6$ self-dual
supergravity, and $\Delta=\ft43$ for the 2-form field strength in $D=5$
simple supergravity.

     It follows from
eqn (\ref{solution1}) that the metrics for the brane solutions are given by
\beq
ds^2 = \Big(1+\fft{k}{r^\dt}\Big)^{-\ft{4\dt}{(D-2)\Delta}}\, dx^\mu dx^\nu
\eta_{\mu\nu} +  \Big(1+\fft{k}{r^\dt}\Big)^{\ft{4d}{(D-2)\Delta}}
\, dy^m dy^m\ .\label{metrsol}
\eeq
This coincides with the results given in ref.\ \cite{dkl} for the
case of $\Delta=4$.  Note from (\ref{betaeq}) and (\ref{solution1}) that in
terms of $\Delta$, the functions $A$, $B$ and $\phi$ satisfy
\beq
A'=\fft{\epsilon\dt}{(D-2)\sqrt\Delta} \, S\ ,\qquad B'=-
\fft{\epsilon d}{(D-2)\sqrt\Delta} \, S\ , \qquad\phi'=\fft{a}{\sqrt\Delta}
\, S\ ,
\label{abphirel}
\eeq
and the dilaton is given by $e^\phi=(1+ kr^{-\tilde d})^{-2a/\Delta
\epsilon}$ with $k=\ft12 \sqrt{\Delta}\lambda/\tilde d$.

       As we shall see in detail in the next section, some of the
$(d-1)$-brane solutions that we have obtained in a $D$-dimensional
supergravity can be isotropically oxidized to $d$-brane solutions in a
$(D+1)$-dimensional supergravity.  The
degree of the antisymmetric tensor involved in a $p$-brane solution, and
the value of the constant $a$, play crucial r\^oles in determining whether
the solution can or cannot be isotropically oxidized in this way.

   At this point, a remark about supersymmetry is in order.  In order for
the solutions that we have obtained above to acquire an interpretation as
{\it super} $(d-1)$-branes embedded in $D$-dimensional spacetime, we shall
have to verify that these solutions preserve half of the supersymmetry of
the corresponding supergravity theories.  We have verified, case by case,
that this is indeed true, at least as long as the antisymmetric tensor is
part of the supergravity multiplet.    In fact, the conditions arising from
the requirement of preserving half of the supersymmetries turn out to be
precisely equivalent to those that we imposed in the ansatz
(\ref{soluans}).

     In concluding this subsection, we return to a more detailed discussion
of a point to which we alluded earlier, namely that we may choose, when
discussing the solution set of elementary and solitonic branes in
supergravity theories, to restrict our attention to the versions of the
various supergravity theories in which all antisymmetric tensors $F_n$ have
degrees $n$ that do not exceed $D/2$.  The reason why we may do this
without losing generality is that an elementary or solitonic solution of a
version of a supergravity theory in which the antisymmetric tensor
participating in the solution is dualised is {\it precisely} the same as
the solitonic or elementary solution, respectively, of the undualised form
of the supergravity theory.  To see this, consider the solitonic solution
of (\ref{eqmo1}), with $F_n$ given by the ansatz (\ref{solans}).  This has
\beq
F_n=\fft{1}{n!} F_{m_1\cdots m_n}\,dy^{m_1}\wedge\cdots\wedge d y^{m_n}=
\fft{\lambda}{n!} e^{-nB}\, \epsilon_{m_1\cdots m_n p}\, \fft{y^p}{r^{n+1}}
\, e^{{\underline m}_1}\wedge \cdots \wedge e^{{\underline m_n}}\ .
\eeq
Thus the Hodge dual of this $n$-form is given by
\beq
*F_n= \fft{\lambda}{(D-n)!}\,  \fft{y^m}{r^{n+1}}\,
e^{-nB}\, \epsilon_{\mu_1\cdots \mu_d}
\, e^{\underline m}\wedge e^{{\underline\mu}_1}\wedge
\cdots \wedge e^{{\underline \mu}_d}\ .\label{fdual}
\eeq
In the dual version of the theory, the $(D-n)$-form
$\widetilde F$ whose Bianchi identity implies the field equation for $F_n$
given in (\ref{eqmo1}) is $\widetilde F= e^{-a\phi} *F_n$, which, from
(\ref{fdual}), has components given by
\beq
{\widetilde F}_{m\mu_1\cdots \mu_d} = \fft{\lambda y^m}{r^{n+1}} \, e^{dA-\dt
B -a\phi}\, \epsilon_{\mu_1\cdots \mu_d}\ .
\eeq
Hence by using (\ref{solution1}), with $\dt=n-1$, we see
that ${\widetilde F}_{m\mu_1\cdots \mu_d}$ is precisely of the form of the
elementary ansatz (\ref{eleans}) for a $(d+1)$-index field strength, where
the function $C$ satisfies its equation of motion (\ref{csol}).  Thus we
see that the solitonic solution of the dualised theory is precisely the
same thing as the elementary solution of the undualised theory, and {\it vice
versa}, with the
antisymmetric tensor written in different variables.  We may therefore,
without loss of generality, consider all supergravity theories in their
versions where the degrees of their antisymmetric tensors $F_n$ satisfy
$n\le D/2$.  The set of all elementary and solitonic brane solutions of these
theories spans the entire set of inequivalent brane solutions of these
theories together with their dualised versions.

\subsection{Kaluza-Klein dimensional reduction}

     In order to describe the processes of oxidation and reduction, we need
to set up the Kaluza-Klein procedure for dimensional reduction from $(D+1)$
to $D$ dimensions.  Let us denote the coordinates of a
$(D+1)$-dimensional spacetime by $x^{\hat\sst{M}}=(x^{\sst{M}}, z)$, where
$z$ is the coordinate of the extra dimension.   The $(D+1)$-dimensional
metric $d\hat s^2$ is related to the $D$-dimensional metric $ds^2$ by
\beq
d\hat s^2 = e^{2\a\varphi} ds^2 + e^{2\beta\varphi} (dz + {\cal
A}_{\sst{M}} dx^{\sst{M}})^2\ ,\label{kkans}
\eeq
where $\varphi$ and ${\cal A}$ are taken to be independent of the extra
coordinate $z$.    The constants $\a$ and $\beta$ will be determined
shortly.  A convenient choice for the vielbein
$\hat e^{\hat\sst{A}}{}_{\hat\sst{M}}$ of the $(D+1)$-dimensional
spacetime is
\bea
\hat e^{\sst{A}}{}_{\sst{M}} = e^{\a\varphi} \, e^{\sst{A}}{}_{\sst{M}}
\ ,&& \hat e^{\underline z}{}_{\sst{M}} = e^{\beta\varphi}\, {\cal
A}_{\sst{M}} \ , \nonumber\\
\hat e^{\sst{A}}{}_{z} = 0\ , && \hat e^{\underline z}{}_z = e^{\beta\varphi}\
{}.
\eea
Note that $M$ and $z$ denote world indices, whilst $A$ and $\underline z$
denote tangent-space indices.

     The spin connection is given by
\bea
\hat \omega^{\sst{AB}} &=& \omega^{\sst{AB}} + \alpha e^{-\alpha\varphi}\,
\Big(\del^{\sst{B}}\varphi\, \hat e^{\sst{A}} - \del^{\sst{A}}\varphi\,
\hat e^{\sst{B}}\Big) -\ft12 {\cal F}^{\sst{AB}} e^{(\beta-2\alpha)\varphi}
\, \hat e^{\underline z}\ ,\nonumber\\ \hat\omega^{{\sst{A}}\underline z}
&=& -\beta e^{-\alpha\varphi}\, \del^{\sst{A}} \varphi\, \hat e^{\underline
z} -\ft12 {\cal F}^{\sst{A}}{}_{\sst{B}} \, e^{(\beta-2\alpha)\varphi}\,
\hat e^{\sst{B}}\ ,\label{spincon}
\eea
where $\del_{\sst{A}}= E_{\sst{A}}{}^{\sst{M}} \del_{\sst{M}}$ is the
partial derivative with a tangent-space index, and ${\cal F}_{\sst{MN}}=
2\del_{[\sst{M}} {\cal A}_{\sst{N}]}$.
Here, $E_{\sst{A}}{}^{\sst{M}}$ is the inverse vielbein in $D$ dimensions.
Choosing $\beta = -(D-2)
\alpha $, we find that the $(D+1)$-dimensional Einstein-Hilbert action
$\hat e\, \hat R$ reduces to
\beq
\hat e  \hat R = e R -(D-1)(D-2) \alpha^2\, e (\del\varphi)^2 -
\ft14 e\, e^{-2(D-1)\alpha\varphi}\, {\cal F}^2\ .\label{ricscal}
\eeq
The Kaluza-Klein dilaton $\varphi$ may be given its canonical normalisation
by choosing the constant $\alpha$ such that
\beq
\alpha^2= \fft{1}{2(D-1)(D-2)}\ .\label{alphaval}
\eeq
It is sometimes useful to have expressions for the $(D+1)$-dimensional
Ricci tensor.  Its tangent-space components are given, after setting
$\beta=-(D-2)\alpha$, by
\bea
\hat R_{\sst{AB}} &=& e^{-2\alpha\varphi} \Big(R_{\sst{AB}} -
(D-1)(D-2)\alpha^2\, \del_{\sst{A}}\varphi\, \del_{\sst{B}}\varphi -
\alpha\,   \square\varphi \, \eta_{\sst{AB}}\Big) -
\ft12 e^{-2D\alpha\varphi}\, {\cal F}_{\sst{A}}{}^{\sst{C}} {\cal F}_{\sst{
BC}}\ ,\nonumber\\
{\hat R}_{{\sst{A}}\underline z} &=& \ft12 e^{(D-3)\a\varphi}\,
\nabla^{\sst{B}}\Big( e^{-2(D-1)\a\varphi}\, {\cal F}_{\sst{AB}} \Big)\ ,
\label{ricten}\\
\hat R_{\underline z \underline z} &=& (D-2)\, \alpha\, e^{-2\alpha\varphi}\,
\square \varphi    +\ft14 e^{-2D\alpha\varphi}\, {\cal F}^2\ .\nonumber
\eea

     Let us now apply the above formalism to the case of a bosonic
Lagrangian of the form (\ref{boslag}), but in $(D+1)$ rather than $D$
dimensions:
\beq
{\cal L} = \hat e \hat R -\ft12 \hat e (\del \hat\phi)^2 - \fft1{2\, n!}
\hat e e^{-\hat a\hat\phi}\,  {\hat F_n}^2\ ,\label{higherbos}
\eeq
where we add a subscript index $n$ to indicate that $F$ is an $n$-form.
The Kaluza-Klein ansatz for $\hat\phi$ is simply $\hat\phi=\phi$, where
$\phi$ is independent of the extra coordinate $z$.  For $\hat F_n$, which
is written locally in terms of a potential $\hat A_{n-1}$ as $\hat F_n=d
\hat A_{n-1}$, the ansatz for $\hat A_{n-1}$ is
\beq
\hat A_{n-1}=B_{n-1} + B_{n-2}\wedge dz\ ,\label{aans}
\eeq
where $B_{n-1}$ and $B_{n-2}$ are potentials for the $n$-form field
strength $G_n=dB_{n-1}$ and the $(n-1)$-form field strength $G_{n-1}=d
B_{n-2}$ in $D$ dimensions.  Defining
\beq
G'_n=G_n - G_{n-1}\wedge {\cal A}\ ,
\eeq
where ${\cal A}={\cal A}_{\sst{M}} dx^{\sst{M}}$, one finds
\beq
\hat F_n = {G'}_n + G_{n-1}\wedge (dz+ {\cal A})\ .\label{transgress}
\eeq
The tangent-space components of $\hat F_n$ in $(D+1)$ dimensions are
therefore given by $\hat F_{\sst{A_1\cdots A_n}}=
{G'}_{\sst{A_1\cdots A_n}} e^{-n\alpha\varphi}$
and $\hat F_{{\sst{A_1\cdots A_{n-1}}} {\underline z}}=
G_{\sst{A_1\cdots A_{n-1}}}
e^{-(n-1)\alpha\varphi -\beta\varphi}$.  Substituting into
(\ref{higherbos}), and using $\beta=-(D-2)\alpha$, we obtain the reduced
$D$-dimensional Lagrangian
\bea
{\cal L} &=& e R -\ft12 e (\del\phi)^2 -\ft12 e (\del\varphi)^2 -
\ft14 e e^{-2(D-1)\alpha\varphi} \, {\cal F}^2 \nonumber\\
&&-\fft{e}{2\, n!}  e^{-2(n-1)\alpha\varphi-\hat a \phi}\, {G'}_n^2 -
\fft{e}{2\, (n-1)!}  e^{2(D-n)\alpha\varphi -\hat a\phi}\,  G_{n-1}^2 \ ,
\label{bosred}
\eea
where $\alpha$ is given by (\ref{alphaval}). As one sees, different
combinations of $\varphi$ and $\phi$ appear in the exponential prefactors
of ${G'}_n^2$ and $G_{n-1}^2$. Nonetheless, each of these prefactors may
easily be seen to be of the form $e^{-a_n\tilde\phi}$, where $\tilde\phi$
is an $SO(2)$ rotated combination of $\varphi$ and $\phi$. In these
prefactors, the coefficients $a_n$ satisfy the formula (\ref{avalue}) in
$D$ dimensions, with $d\dt$ given by (\ref{ndep}), and with the {\it same}
value of $\Delta$ as for $\hat a$ in $(D+1)$ dimensions. (Note that $d\dt$
in (\ref{avalue}) is $n$-dependent, so one obtains different values for the
${G'}_n^2$ and $G_{n-1}^2$ prefactors.)  The 2-form field strength ${\cal
F}$ has an $a$ value given by (\ref{avalue}) with $\Delta=4$.

      Most supergravity theories can be obtained from 11-dimensional
supergravity {\it via} Kaluza-Klein dimensional reduction.   Any such
dimensional reduction can be viewed as a sequence of reductions by one
dimension at a time, of the kind we are discussing here.
Any solution of a lower-dimensional supergravity theory in such a sequence can
therefore be reinterpreted as a solution of any one of the higher
theories in the sequence by use of the Kaluza-Klein ansatz (\ref{kkans}).
In particular, this implies that any elementary or solitonic $p$-brane solution
is also a solution in the higher dimensions.   However, it is important to
realise that the resulting higher dimensional solution may not necessarily
preserve the isotropic form of the $p$-brane ansatz (\ref{metrform}).
In this paper, we are using the term `stainless' to describe the property
of a brane solution of a lower-dimensional supergravity that cannot be
oxidized into an isotropic brane solution in any supergravity in the next
higher dimension.\footnote{We note that in defining a stainless $p$-brane
to be one that cannot be oxidized to an isotropic brane in a higher
dimension, we have not wanted to prejudge what a non-stainless
$p$-brane may oxidize into. {\it A priori}, one could envisage that the
extra dimension acquired upon oxidation could either become isotropically
included into the world-brane dimensions, giving a $(p+1)$-brane in $(D+1)$
dimensions, {\it or} that the extra dimension could be isotropically
included into the transverse dimensions, in which case one would still have
a $p$-brane in the $(D+1)$ dimensions. The latter possibility, however, can
never be realised within the scheme of Kaluza-Klein dimensional reduction
because all fields are by construction taken to be independent of the extra
coordinate, and this would be inconsistent with our ansatz
(\ref{metrform}).} On the other hand, a $(p+1)$-brane solution in $(D+1)$
dimensions necessarily gives rise under dimensional reduction to an
isotropic $p$-brane solution in $D$ dimensions.  This automatic
preservation of isotropicity for solutions under dimensional reduction
corresponds directly to the process of double dimensional reduction
\cite{dhis} of $p$-brane actions.

      The above ideas can be illustrated in our example of the bosonic
Lagrangians (\ref{higherbos}) and (\ref{bosred}).  First, we shall show that
the elementary and solitonic solutions in $(D+1)$ dimensions reduce
respectively to elementary and solitonic solutions in $D$ dimensions.  In
the case of an elementary solution, the $n$-index antisymmetric tensor in
$(D+1)$ dimensions leads to
an elementary brane with world volume dimension $\hat d =n-1$. The elementary
ansatz for the $(D+1)$-dimensional field strength $\hat F_n$ in
(\ref{higherbos}) is $\hat F_{m\mu_1\ldots\mu_{n-2} z} =
\epsilon_{\mu_1\ldots\mu_{n-2} z} \del_m e^C$.  It follows from
eqn (\ref{transgress}) that the corresponding $D$ dimensional fields
$G_{n-1}$, ${G'}_n$ and ${\cal A}$ become
\bea
G_{m\mu_1\ldots\mu_{n-2}} &=& \epsilon_{\mu_1\ldots\mu_{n-2}} \del_m e^C
\ ,\nonumber\\
{G'}_{\sst{M_1}\ldots\sst{M_n}} &=& 0\ ,\qquad
{\cal A}_{\sst{M}}=0 \ .\label{elered}
\eea
This is nothing but the usual elementary-type ansatz for an $(n-1)$-index
antisymmetric tensor in $D$ dimensions, and thus gives rise to an elementary
brane solution (\ref{solution1}) with world volume dimension $d=n-2$.

      The metric ansatz in $(D+1)$ dimensions is given by $d\hat s^2 =
e^{2\hat A}(dx^\mu dx^\nu \eta_{\mu\nu} + dz^2) + e^{2\hat B} dy^m dy^m$.
In the elementary solution in $(D+1)$ dimensions, it follows from
(\ref{solution1}) that $\phi= \hat a (D-1) \hat A/\dt$, and $\hat B= -(d+1)
\hat A/\dt$.  (Note that $\dt$ is the same for both $D$ and $(D+1)$
dimensions since, by definition, $\dt +2$ is the codimension of the world
volume of the brane.)   On the other hand in $D$ dimensions, we see from
(\ref{bosred}) that the combination of scalar fields $-2(D-n)\a\varphi+
\hat a\phi=a\tilde\phi$, with $a^2=\hat a^2 +4(D-n)^2 \a^2$, defines the
$SO(2)$-rotated $D$-dimensional dilaton $\tilde \phi$, whilst the
orthogonal combination $2(D-n)\a \phi+ \hat a\varphi$ is set to zero.
Since $n=d+2$, it then follows from (\ref{alphaval}) that $\hat a$
and $a$ are related by
\beq
\hat a^2 = a^2 - \fft{2\dt^2}{(D-2)(D-1)}\ .\label{arelation}
\eeq
Thus we find that $\tilde\phi=a (D-2) A/\dt$, $B=-d A/\dt$ and $e^{c A}=
e^{\hat c\hat A}$, since, from
the Kaluza-Klein ansatz (\ref{kkans}) for the metric, we have $\hat
A=A+\a\varphi$ and $\hat B= B +\a\varphi$.  But these expressions for
$\tilde\phi$ and $B$ are precisely of the form given in (\ref{solution1})
for the elementary $(d-1)$-brane in $D$ dimensions.  Thus we conclude that
under dimensional reduction, an elementary $d$-brane in $(D+1)$ reduces to
an elementary $(d-1)$-brane in $D$ dimensions.

     In the case of solitonic solutions, the analysis is parallel.  The
ansatz for the $n$-index antisymmetric tensor, which leads to a solitonic
brane solution with world volume dimension $d=D-n$ in $(D+1)$ dimensions,
takes the form $\hat F_{m_1\ldots m_n} = \lambda \epsilon_{m_1\ldots m_n p}
\, y^p\, r^{-n-1}$. It follows from eqn (\ref{transgress}) that the
corresponding $D$ dimensional fields ${G'}_n$, $G_{n-1}$ and ${\cal A}$ become
\bea
{G'}_{m_1\ldots m_n} &=&\lambda \epsilon_{m_1\ldots m_n
p}\, y^p\, r^{-n-1} \ ,\nonumber\\
G_{\sst{M_1}\ldots\sst{M_{n-1}}} &=& 0\ ,\qquad
{\cal A}_{\sst{M}} = 0 \ .
\eea
This is indeed just the field configuration for a solitonic $(d-1)$-brane
in $D$ dimensions.   The analysis of the relation between the metrics in
$(D+1)$ and $D$ dimensions is very similar to that in the elementary case.

       It is of interest to note that in the reduction of a $d$-brane in
$(D+1)$ dimensions to a $(d-1)$ brane in $D$ dimensions, the degree of the
antisymmetric tensor involved in the solution reduces by one in the
elementary case, but remains unchanged in the solitonic case.   Note also
that the relation between $\hat a$ and $a$ in eqn (\ref{arelation}) is
always satisfied in the dimensional reduction of a brane solution in $(D+1)$
to one in $D$ dimensions.  This implies, conversely, that  eqn
(\ref{arelation}) is a {\it necessary} condition for the reverse procedure
to be possible.  It is easy to verify that the relation (\ref{arelation})
is uniquely satisfied with $\hat a$ and $a$ given by eqn (\ref{avalue}),
provided that $\Delta$ is the same for both $\hat a$ and $a$.

     We have seen that brane solutions in higher dimensions can be reduced
to those in lower dimensions {\it via} the Kaluza-Klein procedure; however,
the inverse procedure is not necessarily possible.  For example the
$D$-dimensional bosonic Lagrangian (\ref{bosred}) that is derived from the
$(D+1)$-dimensional Lagrangian (\ref{higherbos}) admits six brane
solutions, namely an elementary and a solitonic solution for each of the
three antisymmetric tensors $G_n$, $ G_{n-1}$ and $\cal F$.  Two of these
solutions are isotropically oxidizable to brane solutions in $(D+1)$
dimensions, by reversing the procedure discussed above, namely the elementary
solution using $G_{n-1}$ and the solitonic solution using $G_n$. The remaining
four solutions are stainless because they cannot be oxidized to isotropic brane
solutions of the $(D+1)$ dimensional theory defined by eqn (\ref{higherbos}).
To illustrate this, consider the elementary solution that uses the
antisymmetric tensor $G_n$ in the $D$-dimensional Lagrangian (\ref{bosred}).
The solution for the metric in $D$ dimensions is given by (\ref{metrform}) with
$A$ and $B$ given in eqn (\ref{solution1}).   This solution can be oxidized
into a solution in $(D+1)$ dimensions, whose metric is given by
\beq
d\hat s^2 = e^{2\hat A} dx^\mu dx^\nu\eta_{\mu\nu} +
e^{2\hat B} (dy^m dy^m + dz^2)\ ,\label{genoxiele}
\eeq
where
\beq
\hat A = \fft{(D-2)(\dt +1)}{(D-1)\dt} \, A\ ,\qquad
\hat B = - \fft{(D-2)d}{(D-1)\dt}\,  A\ .\label{noname2}
\eeq
{}From the form of this $(D+1)$-dimensional metric, we can see that it does
not describe an isotropic $d$-brane, since the different $r$-dependent
prefactor for $dz^2$ prevents $z$ from being grouped together with the
coordinates $x^\mu$.   Note also that, although $dz^2$ does have the same
prefactor as $dy^mdy^m$, this metric is still not isotropic in the
transverse directions because the prefactors $e^{2\hat A}$ and $e^{2\hat
B}$ are functions of $r=\sqrt{y^my^m}$ and not of $\sqrt{y^my^m + z^2}$.

      To summarise, we have seen that an elementary or solitonic
$(p+1)$-brane solution in $(D+1)$ dimensions can always be reduced respectively
to an elementary or solitonic $p$-brane solution in $D$ dimensions.  On the
other hand, the inverse process of dimensional oxidation to an isotropic
brane solution is not always possible.  Thus in a brane scan of elementary
and solitonic solutions, we may factor out the rusty solutions and
characterise the full solution set by the stainless $p$-branes only.

       There are three cases in which a $p$-brane solution
can turn out to be stainless.  The first case is when a brane solution
arises in a supergravity theory that cannot be obtained by dimensional
reduction, such as $D=11$ supergravity or type IIB supergravity in $D=10$.
In the remaining two cases, the supergravity theory itself can be obtained
by dimensional reduction, but oxidation to an isotropic brane solution is
nonetheless not possible.   In the second case, no $(D+1)$-dimensional
supergravity theory has the necessary antisymmetric tensor for an isotropic
brane solution.   Specifically, if the $D$-dimensional solution is
elementary, the $(D+1)$-dimensional theory would need an antisymmetric
tensor of degree one higher than that in the $D$-dimensional theory. If it
is instead a solitonic solution, the $(D+1)$-dimensional theory would need
an antisymmetric tensor of the same degree as in the $D$-dimensional
theory.  In the third case, an antisymmetric tensor of the required degree
exists in the $(D+1)$-dimensional theory, but the exponential dilaton
prefactor has a coefficient $\hat a$ that does not satisfy eqn
(\ref{arelation}).   We shall meet examples of all three cases in the
subsequent sections.

\section{$D\ge10$ supergravity}

      $D=11$ is the highest dimension for any supergravity theory, and
hence all the $D=11$ $p$-brane solutions are necessarily stainless. Since
there is only one antisymmetric tensor field strength in the theory, namely
a 4-index field, there is just one elementary membrane solution \cite{dust} and
one solitonic 5-brane solution \cite{guv}. (Original papers giving $D=11$
supergravity, and all the other supergravities in various dimensions that we
will consider here, can be found in \cite{SS}.)

     Dimensional reduction of $D=11$ supergravity to $D=10$ yields type
IIA supergravity.  The type IIA theory contains: a 2-form field strength
giving rise to a particle and a 6-brane; a 3-form giving rise to a string
and a 5-brane; and a 4-form giving rise to a membrane and a 4-brane.  In
each case we have listed first the elementary and then the solitonic
solution.   All of these solutions break half of the $D=10$, $N=2$
supersymmetry.  Of the six solutions two, namely the elementary string
and the solitonic 4-brane, can be oxidized to the corresponding elementary
membrane and solitonic 5-brane in $D=11$.   The remaining four solutions
are stainless since $D=11$ supergravity lacks the necessary antisymmetric
tensors.   Note that the $11\longrightarrow 10$ situation corresponds
precisely to the bosonic example we discussed in section 2.2.

     In addition, in $D=10$, there is the type IIB supergravity, which
cannot be obtained by dimensional reduction from $D=11$.  This theory
contains a complex 3-form field strength giving rise to an elementary
string and a solitonic 5-brane solution; and a self-dual 5-form
field strength giving rise to a self-dual 3-brane \cite{dulu}. The string and
5-brane are in fact also solutions of $D=10$, $N=1$ supergravity, and are hence
identical to the string and 5-brane solutions of the type IIA theory.
Thus, although the type IIB theory cannot itself be obtained by
dimensional reduction from $D=11$, these particular solutions of the IIB
theory do have an oxidation pathway up to isotropic solutions in $D=11$.
In such situations, we do not consider brane solutions to be stainless.
The remaining solution, the self-dual 3-brane, is the only solution that
belongs exclusively to the IIB theory. It is stainless and breaks half of
the $N=2$ supersymmetry.

\section{$D=9$ supergravity}

\subsection{$N=1$, $D=9$ supergravity}

      $N=1$ supergravity in $D=9$ \cite{sgn} contains a 2-form field strength
giving rise to an elementary particle and a solitonic 5-brane; and a 3-form
field strength giving rise to an elementary string and a solitonic
4-brane. The solitonic 4-brane solution can be isotropically oxidized to the
solitonic 5-brane of $N=1$, $D=10$ supergravity. The situation is somewhat
more complicated for the oxidation of the elementary string solution.
Obviously, this solution cannot be oxidized isotropically to an elementary
membrane solution of $N=1$, $D=10$ supergravity because this theory lacks
the necessary 4-form field strength, and thus no elementary membrane exists
in the $N=1$, $D=10$ theory.  Nonetheless, the $D=9$ string solution is not
stainless because there is a different oxidation pathway available to it.
The $D=9$ string can also be viewed as a solution of $N=2$, $D=9$
supergravity. In this guise, it can oxidize isotropically to a solution of
type IIA $D=10$ supergravity, which {\it does} have a 4-form field strength.

     The elementary particle and solitonic 5-brane solution that arise
from the 2-form field strength are stainless. Na\"\i vely, one might
expect these solutions could oxidize up to the elementary string and
solitonic 6-brane solutions of type IIA $D=10$ supergravity. However, as we
showed in section 2.2, even when the necessary forms are present in the
higher-dimensional theory an isotropic oxidation is possible only when
the coefficient $a$ appearing in the dilaton prefactor $e^{-a\phi}$
satisfies the relation (\ref{arelation}). In the case of $N=1$, $D=9$
supergravity, the coefficient $a$ is given by eqn (\ref{avalue}) with
$\Delta=2$. On the other hand, the coefficient $a$ in the type IIA, $D=10$
theory is given by eqn (\ref{avalue}) with $\Delta=4$. Since the $\Delta$
value has to be preserved under dimensional reduction, it follows that the
particle and 5-brane solutions in $D=9$ are stainless.

     There {\it are} elementary particle and solitonic 5-brane
descendants in $D=9$, nonetheless. These {\it are} obtained by dimensional
reduction from the type IIA $D=10$ elementary membrane and solitonic
6-brane. From the $D=9$ point of view, these are obtained as solutions to
$N=2$ supergravity using a 2-form field strength whose dilaton prefactor
indeed has an $a$ coefficient given by (\ref{avalue}) with the necessary
$\Delta=4$. The difference in $\Delta$ values establishes the distinctness
of the stainless particle and 5-brane discussed above from those obtained
by dimensional reduction.   The metrics for the stainless elementary
particle and solitonic 5-brane are given by
\bea
{\rm elementary}:&& ds^2 = - \Big(1+\fft{k}{r^{6}}\Big)^{-12/7} dt^2  +
\Big(1+\fft{k}{r^{6}}\Big)^{2/7} dy^mdy^m \ ,\nonumber\\
{\rm solitonic}:&& ds^2 =\Big (1+\fft{k}{r}\Big)^{-2/7} dx^\mu dx^\nu
\eta_{\mu\nu} + \Big(1+\fft{k}{r}\Big)^{12/7} dy^mdy^m\ .
\eea
By contrast, the metrics for the elementary particle and solitonic
5-brane that can oxidize to an elementary string and a solitonic 6-brane
in $D=10$ are given by
\bea
{\rm elementary}:&& ds^2 = - \Big(1+\fft{k}{r^{6}}\Big)^{-6/7} dt^2  +
\Big(1+\fft{k}{r^{6}}\Big)^{1/7} dy^mdy^m \ ,\nonumber\\
{\rm solitonic}:&& ds^2 = \Big(1+\fft{k}{r}\Big)^{-1/7} dx^\mu dx^\nu
\eta_{\mu\nu} + \Big(1+\fft{k}{r}\Big)^{6/7} dy^mdy^m\ .
\eea

     Let us now examine in detail the new stainless $D=9$ solutions. In
particular, we need to verify that they preserve half of the
supersymmetry. Since these solutions cannot be obtained from isotropic
solutions in $D=10$, we
do not have an automatic guarantee that half of the supersymmetry will be
preserved. To investigate this, we first give the bosonic sector of the
Lagrangian and the supersymmetry transformations. The bosonic sector of
the Lagrangian is
\beq
{\cal L}=eR-\ft12e(\del\phi)^2-
\ft1{12}ee^{-\sqrt{\fft87}\phi}\, G_{\sst{MNP}}G^{\sst{MNP}}-
\ft14ee^{-\sqrt{\fft27}\phi}\, F_{\sst{MN}}F^{\sst{MN}}\ ,
\label{d9lag}
\eeq
where $F_{\sst{MN}}=2\del_{[\sst M}A_{\sst N]}$ and
$G_{\sst{MNP}}=3\del_{[\sst M}B_{\sst{NP}]}+\ft32 A_{[\sst M}F_{\sst{NP}]}$
\cite{sgn}. By
comparison with eqn (\ref{avalue}) it is easy to verify that the
$\Delta$ value for the 3-form $G$ is 4, but the value for the 2-form $F$
is 2. The supersymmetry transformation rules for the bosonic fields are:
\bea
\delta e^{\sst A}{}_{\sst M} &=& -\im \, \bar\varepsilon\Gamma^{\sst
A}\psi_{\sst M}\ ,\qquad \delta\phi = \im\sqrt2\, \bar\varepsilon\chi\
,\nonumber\\
\delta A_{\sst M} &=&
-\ft2{\sqrt{14}}\, e^{\sqrt{\fft1{14}}\phi}\,
\bar\varepsilon\Gamma_{\sst M}\chi
+ \sqrt2 \, e^{\sqrt{\fft1{14}}\phi}\, \bar\varepsilon\psi_{\sst M}\ ,
\label{d9bostr}\\
\delta B_{\sst{MN}} &=& -2\im\,
e^{\sqrt{\fft27}\phi}\, \bar\varepsilon\Gamma_{[\sst M}\psi_{\sst N]} +
\ft{2\im}{\sqrt{7}}\, e^{\sqrt{\fft27}\phi}\, \bar\varepsilon
\Gamma_{\sst{MN}}\chi +
A_{[\sst M}\delta A_{\sst N]}\ .\nonumber
\eea
For the fermionic fields, the supersymmetry transformations are:
\bea
\delta\chi &=& -\ft1{2\sqrt2}\Gamma^{\sst M}\varepsilon\, \del_{\sst M}\phi
+\ft1{12\sqrt7}\, e^{-\sqrt{\fft27}\phi}\, G_{\sst{MNP}}
\Gamma^{\sst{MNP}}\varepsilon
-\ft{\im}{4\sqrt{14}}\, e^{-\sqrt{\fft1{14}}\phi}\, F_{\sst{MN}}
\Gamma^{\sst{MN}}\varepsilon\ ,\nonumber\\
\delta\psi_{\sst M} &=& D_{\sst M}\varepsilon +
\ft1{84}e^{-\sqrt{\fft27}\phi}\, G_{\sst{NPQ}}\Big(
\Gamma_{\sst M}{}^{\sst{NPQ}}
- \ft{15}2\delta_{\sst M}^{\sst N}\, \Gamma^{\sst{PQ}}\Big)\varepsilon
\nonumber\\
&& -\ft{\im}{28\sqrt2}\, e^{-\sqrt{\fft1{14}}\phi}\, F_{\sst{NP}}\Big(
\Gamma_{\sst M}{}^{\sst{NP}} - 12\delta_{\sst M}^{\sst N}\, \Gamma^{\sst
P}\Big)\varepsilon\ .\label{d9fertr}
\eea

     The elementary particle and solitonic 5-brane in $D=9$ dimensions are
obtained from the ans\"atze for the 2-index antisymmetric tensor field
strength $F_{\sst{MN}}$ given in (\ref{eleans}) and (\ref{solans})
respectively.  The solutions are given by (\ref{solution1}).  We shall
first verify that the solitonic 5-brane solution preserves half of the
supersymmetry.  We begin by making a $6+3$ split of the gamma matrices:
\beq
\Gamma^{\mu} = \gamma^\mu \otimes \oneone\ ,\qquad
\Gamma^{m} = \gamma_7 \otimes \gamma^m \ ,\label{d95bs}
\eeq
where $\gamma_7 = \gamma_0\gamma_1\ldots\gamma_5$ on the world volume and
$\gamma_1\gamma_2\gamma_3=\im$ in the transverse space.  Here, and
throughout the paper, we adopt the convention that $\gamma_\mu$ and
$\gamma_m$ are purely numerical matrices, with flat indices.
The transformation rules for the fermionic fields in (\ref{d9fertr}) become
\bea
\delta \chi &=& -\fft1{2\sqrt{2}}\, e^{-B}\, \del_m \phi\, \gamma_7\otimes
\gamma_m \varepsilon + \fft{\lambda}{2\sqrt{14}}\,
e^{ -2B-\sqrt{\fft1{14}}\phi
}\, \fft{y^m}{r^3}\, \oneone \otimes \gamma_m \varepsilon \ ,\nonumber\\
\delta \psi_\mu&=& \fft12 \del_m A\, e^{A-B}\, \gamma_\mu \gamma_7\otimes
\gamma_m \varepsilon + \fft\lambda{14\sqrt2}\, e^{A-2B-\sqrt{\fft1{14}}\phi}
\, \fft{y^m}{r^3}\, \gamma_\mu\otimes \gamma_m \varepsilon\ ,\label{d9soltr}\\
\delta \psi_m &=& \del_m \varepsilon + \fft{\im}2 \del_n B
\varepsilon_{mnp} 1\otimes \gamma_p +
\fft{\lambda}{14\sqrt2}\, e^{-B-\sqrt{\fft1{14}}\phi}\, \fft{y^m}{r^3}\,
\gamma_7\otimes \oneone\,  \varepsilon \nonumber\\
 &&- \fft{3\im\lambda}{7\sqrt{2}}\,
e^{- B-\sqrt{\fft1{14}}\phi} \, \varepsilon_{mnp} \,
\fft{y^n}{r^3}\, \gamma_7\otimes \gamma_p\, \varepsilon\ .\nonumber
\eea
Substituting the solitonic solution (\ref{solution1}) and noting that
$\phi'$ and $A'$ satisfy (\ref{soluans}) and (\ref{betaeq}), we find that
these variations all vanish provided that
\beq
\varepsilon = e^{\ft12 A} \, \varepsilon_0\ ,\qquad
\gamma_7\otimes\oneone\, \varepsilon_0 = \varepsilon_0\ ,\label{d9epso}
\eeq
where $\varepsilon_0$ is a constant spinor.  Thus our solitonic 5-brane
solution preserves half of the supersymmetry.

     We shall now verify that the elementary particle solution also
preserves half the supersymmetry.   We make a $1+8$ split of the gamma
matrices:
\beq
\Gamma^0 = \im \gamma_9\ ,\qquad \Gamma^m = \gamma^m\ ,
\eeq
where $\gamma_9=\gamma_1\gamma_2\cdots\gamma_8$.  The transformation rules
(\ref{d9fertr}) for the fermionic fields become
\bea
\delta \chi &=& -\ft1{2\sqrt2}\, e^{-B} \del_m \phi\, \gamma_m \varepsilon
+ \ft1{2\sqrt{14}}\, e^{-A - B +C- \sqrt{\fft1{14}}\phi}\, \del_m C \,\gamma_m
\gamma_9 \varepsilon\ ,\nonumber\\
\delta \psi_0 &=& \ft\im{2} e^{A-B}\, \del_m A\, \gamma_m\gamma_9 \varepsilon
-\ft{3\im}{7\sqrt2}\, e^{-B+C-\sqrt{\fft1{14}}\phi}\, \del_m C
\,\gamma_m \varepsilon\ ,\label{d9eletr}\\
\delta \psi_m &=& \del_m \varepsilon +\ft{\im}2 \del_n B\, \gamma_{mn}
\varepsilon + \ft{1}{14\sqrt2}\, e^{-A +C-\sqrt{\fft1{14}}\phi}\, \del_m C\,
\gamma_{mn}\gamma_9 \varepsilon\nonumber\\
&& -\ft3{7\sqrt{2}}\, e^{-A+C  -\sqrt{\fft1{14}}\phi} \,
\del_m C \,\gamma_9 \varepsilon\ .\nonumber
\eea
Analogously to the solitonic case, the elementary particle solution also
preserves half of the supersymmetry provided that
\beq
\varepsilon = e^{\ft12 A}\, \varepsilon_0\ ,\qquad
\gamma_9\,  \varepsilon_0  =\varepsilon_0\ ,\label{d9epel}
\eeq

     So far we have obtained a stainless elementary particle and stainless
solitonic 5-brane.  Both solutions break half of the supersymmetry.  The
reason why these two solutions cannot be isotropically oxidized into
$D=10$ dimensions is that both are obtained from the 2-index
antisymmetric tensor field strength with the dilaton prefactor
$e^{-a\phi}$ where the $a$ coefficient is given by (\ref{avalue}) with
$\Delta = 2$, instead of the value $\Delta=4$ that characterises the
prefactors of antisymmetric tensor field strengths in $D=10$.
At first sight the occurrence of this new value of $\Delta$
may seem paradoxical since, Kaluza-Klein dimensional reduction preserves
the value of $\Delta$, as we discussed for the scalar field $\tilde \phi$
defined below eqn (\ref{bosred}). Since $N=1$, $D=9$ supergravity can be
obtained by dimensional reduction of $N=1$, $D=10$ supergravity, which has
a single 3-form field strength, with a $\Delta = 4$ prefactor, it follows
that all the antisymmetric tensors in $D=9$ will have $\Delta=4$
prefactors.

     The resolution of this apparent paradox involves details of the
truncation of dimensionally reduced $N=1$, $D=10$ supergravity to the
pure $N=1$ supergravity multiplet in $D=9$.  The truncation removes a
single $D=9$ Maxwell multiplet.   The Lagrangian of the bosonic sector of
$N=1$, $D=10$ supergravity is
\beq
{\cal L} = \hat e \hat R - \ft12 e (\del\phi)^2 -\ft1{12} \hat e e^{-\phi}
\, \hat F_{\sst{MNP}} \hat F^{\sst{MNP}}\ .
\eeq
Following the Kaluza-Klein dimensional reduction scheme discussed in
section 2.2, this leads to the $D=9$ Lagrangian
\bea
{\cal L} &=& e R -\ft12 e(\del\phi)^2 -\ft12 e(\del\varphi)^2 -\ft1{12} e
e^{-\phi -\sqrt{\fft17}\varphi}\, {G'}_3^2 \nonumber\\
&& -\ft14 e e^{-\fft4{\sqrt7}\varphi}\, {\cal F}^2 -\ft14 e e^{-\phi +
\fft3{\sqrt7}\varphi}\, G_2^2\ .\label{n1d9lag}
\eea
As it stands, one cannot consistently truncate out either of the 2-form
field strengths or either of the two scalars.   Nonetheless, it is possible
to make a consistent truncation to the bosonic sector of pure $N=1$, $D=9$
supergravity.\footnote{The possibility of making a consistent truncation to
the $N=1$, $D=9$ supermultiplet may be shown using arguments similar to
those in ref.\ \cite{zilch}}  In order to do this, we must first rotate the
basis for the scalar fields:
\beq
\phi = \sqrt{\ft78} \phi_1 -\sqrt{\ft18} \phi_2\ ,\qquad
\varphi = \sqrt{\ft18} \phi_1 + \sqrt{\ft78} \phi_2\ .
\eeq
In terms of this rotated basis, the Lagrangian (\ref{n1d9lag}) becomes
\bea
{\cal L} &=& e R -\ft12 e(\del\phi_1)^2 -\ft12 e(\del\phi_2)^2 -\ft1{12} e
e^{-\sqrt{\fft87} \phi_1 }\, {G'}_3^2 \nonumber\\
&& -\ft14 e e^{-\sqrt{\fft27} \phi_1 -\sqrt{2} \phi_2 }\, {\cal F}^2 -\ft14 e
e^{- \sqrt{\fft27} \phi_1 +\sqrt{2}\phi_2}\, G_2^2\ .\label{n1d9lag2}
\eea
Now we can see that it is consistent with the equation of motion for
$\phi_2$ to set $\phi_2=0$ provided that at the same time we set ${\cal F}$
equal to $G_2$.  Defining then $F= \sqrt2{\cal F} = \sqrt2 G_2$, we obtain the
Lagrangian for the bosonic sector of pure $N=1$, $D=9$ supergravity:
\beq
{\cal L} = e R -\ft12 e(\del \phi_1)^2 -\ft1{12} e e^{-\sqrt{\fft87}\phi_1}
{G'}_3^2 -\ft14 e e^{-\sqrt{\fft27}\phi_1} F^2\ ,\label{n1d9sez}
\eeq
where $d {G'}_3 + \ft12 F\wedge F =0$.  This result coincides with the
Lagrangian given in ref.\ \cite{sgn}, which appears in eqn (\ref{d9lag}).
Thus although the value of $\Delta$ for the combinations $- \sqrt{\ft27}
\phi_1 \pm \sqrt{2}\phi_2$ occurring in the 2-form field strength
prefactors before truncation is $\Delta=4$, the value after the truncation
in which $\phi_2$ is set equal to zero is $\Delta = 2 $.\footnote{
Another point of view for resolving the apparent paradox is to
regard the stainless particle and 5-brane as solutions of the full
dimensionally reduced $N=1$, $D=10$ supergravity, {\it i.e.}\ $N=1$, $D=9$
supergravity plus the Maxwell multiplet.  From this point of view, these
solutions fall outside out $p$-brane ans\"atze (\ref{eleans}) and
(\ref{solans}) because more than one antisymmetric tensor field strength
takes a non-vanishing value.  The solutions arising from this new ansatz
are equivalent to those in the truncated $N=1$ theory with $\Delta = 2$.}

     Having studied this example in detail, we are now in a position to be
more precise about the possible values of $\Delta$ that can arise in
supergravity theories.  We have seen that we may treat $D=11$
supergravity, which has no dilaton, as having the value $\Delta=4$ for its
4-form field strength, since this value corresponds, by virtue of eqn
(\ref{avalue}), to $a=0$.  We have also seen that pure Kaluza-Klein
dimensional reduction, where one performs no truncation on the
lower-dimensional theory, preserves the values of $\Delta$ from the higher
dimension.  Thus in the absence of any truncation, all supergravity theories
that are obtained by dimensional reduction from $D=11$ will have $\Delta=4$
for all dilaton couplings.  However, as we demonstrated in the case of
$N=1$, $D=9$  supergravity above, if a supergravity theory in a
lower dimension is
obtained by a process of {\it truncation} as well as dimensional reduction,
then the values of $\Delta$ for the coupling of the particular combinations of
dilaton fields that survive the truncation to the antisymmetric tensor
combinations that survive the truncation can differ from 4.  For example,
one can have $\Delta=2$ for 2-form field strengths in $D\le9$
supergravities.

     Before ending this section, it is of interest to investigate the warped
metrics that one does obtain in $D=10$ if one oxidizes the stainless
elementary particle and solitonic 5-brane from $D=9$, so as to compare them
with the isotropic metrics of the elementary string and and solitonic
6-brane occurring in $D=10$. The metrics obtained by oxidizing the stainless
$D=9$ solutions are given by
\bea
{\rm elementary:}&& d\hat s^2 = \Big ( 1 + \fft{k}{r^6}\Big )^{-7/4} dx^\mu
dx^\nu\eta_{\mu\nu} + \Big (1+ \fft{k}{r^6} \Big)^{1/4} \Big (dy^m dy^m +
(dz + {\cal A})^2\Big)\ , \nonumber\\
{\rm solitonic:} && d\hat s^2 = \Big ( 1 + \fft{k}{r}\Big )^{-1/4}
\Big( dx^\mu dx^\nu\eta_{\mu\nu} + (dz + {\cal A})^2\Big) +
\Big (1+ \fft{k}{r} \Big)^{7/4} dy^m dy^m\ .\label{d9oxid}
\eea
Here we see that we have pushed oxidation too far: neither of these two
metrics describes isotropic brane solutions in $D=10$.  In both cases there
is a non-vanishing gauge potential ${\cal A}={\cal A}_{\sst M} dx^{\sst M}$,
which describes a
topologically non-trivial field configuration, implying that $z$ is a
coordinate on a non-trivial $U(1)$ fibre bundle, and thus the metric is
`twisted.' Furthermore, in order for this coordinate to be well defined, it
must be taken to be periodic with period $\Delta z = \int {\cal F}$ (or
$\int{\cal F}$ divided by any integer).  In the elementary case, as we also
saw in the general example given in eqn (\ref{genoxiele}),  the metric would
not be isotropic even if ${\cal A}$ were equal to zero, for the reasons we
discussed.   By contrast, the metrics for the
{\it isotropic} elementary string and solitonic 6-brane are given by
(\ref{metrsol}) with $D=10$ and $\Delta = 4$, by taking $d=2$ and $d=7$
respectively.

\subsection{$N=2$, $D=9$ supergravity}

     $N=2$ supergravity in $D=9$ contains three 2-form, two 3-form and one
4-form field strengths.   In addition there are three scalar fields.  Two
of these behave like dilatons and appear undifferentiated in exponential
prefactors multiplying the kinetic terms for the antisymmetric tensors. The
third scalar does not appear in exponential prefactors in the Lagrangian;
furthermore, its kinetic term itself has a dilaton prefactor.  Thus we may
view this scalar field as the 0-form potential for a 1-form field strength.
We can use this field strength to obtain a solitonic 6-brane in $D=9$.

    $N=2$, $D=9$ supergravity has not yet been constructed; however, it
could be easily obtained by dimensional reduction of type IIA supergravity in
$D=10$.   We expect that the elementary and solitonic brane solutions that
are obtained from the 2-form, 3-form and 4-form field strengths are either
obtainable by dimensional reduction from those in $D=10$ or are
equivalent to the stainless solutions we constructed in $N=1$, $D=9$
supergravity.  However, the solitonic 6-brane that is associated with the
1-form field strength is necessarily stainless, since the 1-form field
strength appears first in $D=9$ supergravity in the descent from eleven
dimensions.  We shall first obtain the solution and then shall verify that it
preserves half of the supersymmetry.

    The Lagrangian of the relevant part of the bosonic sector of $N=2$, $D=9$
supergravity can be obtained by Kaluza-Klein dimensional reduction of the
metric, dilaton and 2-form field strength in type IIA supergravity in
$D=10$, whose Lagrangian is given by eqn (\ref{higherbos}) with $n=2$ and
$\hat a = 3/2$.  The reduced $9$-dimensional Lagrangian is given by
(\ref{bosred}), again with $n=2$ and $\hat a=3/2$.  In order to obtain a
solitonic 6-brane solution, we can consistently set ${\cal F}=0$,
${G'}_2=0$ and furthermore truncate out one of the two scalar field degrees of
freedom by setting
\beq
\ft32 \phi - 14\a \varphi = 2 \tilde \phi\ ,\qquad
14\a \phi + \ft32 \varphi = 0\ .
\eeq
Thus the Lagrangian for the relevant bosonic fields in $D=9$ is
\beq
{\cal L}  = e R - \ft12 e (\del \tilde\phi)^2 - \ft12 e  e^{-2\tilde \phi}
\, G_1^2\ .
\eeq
This construction, which precisely parallels the previous discussions for
general values of $n$, emphasises that $G_{\sst M}=\del_{\sst{M}}b$ should
properly be thought of as the field strength for the 0-form gauge potential
$b=\hat A_z$, since it has its origin in the gauge field $\hat F_2$ in
$D=10$.  Thus it is legitimate for $G_1$ to take the necessary
topologically non-trivial form in the solitonic 6-brane solution in $D=9$,
in which its 0-form potential is well-defined only in patches.

     Using the ansatz for $G_1$ given by eqn
(\ref{solans}), we can obtain the 6-brane solution.  However in this case
$\dt=0$, and hence the general solution given by eqn (\ref{solution1}) no
longer applies.   Nonetheless the equations (\ref{eqmo2}) are easy to
solve; the metric of the solitonic 6-brane in $D=9$ is given by
\beq
ds^2 = dx^\mu dx^\nu \eta_{\mu\nu} + (1 + k \log{r}) \, dy^m dy^m\ ,
\eeq
and the dilaton field $\tilde \phi$ is given by $e^{\tilde \phi} = 1+ k
\log r$.  It satisfies $\tilde\phi'=S$, where $S$ is given in eqn
(\ref{sforms}).

     If $N=2$, $D=9$ supergravity had been constructed, it would have been a
simple matter to check whether the above solution preserved half of the
supersymmetry.   In lieu of this, we may exploit the fact that Kaluza-Klein
dimensional reduction preserves unbroken supersymmetry, and carry out the
computation for the corresponding oxidized brane solution in $D=10$.    Of
course, since the 6-brane in $D=9$ is stainless, the resulting oxidized
metric will not be an isotropic 7-brane.  In fact it takes the form
\beq
d\hat s^2 = e^{-\fft18 \tilde\phi}\, dx^\mu dx^\nu \eta_{\mu\nu} +
e^{\fft78 \tilde\phi}\, \Big (dy^m dy^m + dz^2\Big)\ .\label{ox6brane}
\eeq
The relevant terms in the fermionic transformation rules of type IIA,
$D=10$ supergravity, involving the non-vanishing
2-form field strength $\hat F_2$, are:
\bea
\delta \chi &=& \ft{\sqrt2}{4} \del_{\hat\sst{M}}\phi\, \hat\Gamma^{\hat
\sst{M}} \hat\Gamma^{11}\varepsilon -\ft{3}{16\sqrt 2} e^{-\fft34\phi}\,
\hat F_{\hat\sst{M}\hat\sst{N}} \hat\Gamma^{\hat\sst{M}\hat\sst{N}}
\varepsilon\ ,\nonumber\\
\delta \psi_{\hat\sst{M}} &=& \hat D_{\hat\sst{M}}\varepsilon -\ft1{64}
e^{-\fft34 \phi}\, \hat F_{\hat\sst{N}\hat\sst{P}}\Big(
\hat\Gamma_{\hat\sst{M}}{}^{
\hat\sst{N}\hat\sst{P}} -14 \delta_{\hat\sst{M}}^{\hat\sst{N}}\,
\hat\Gamma^{\hat\sst{P}}\Big)\hat\Gamma^{11}\varepsilon \ .\label{d10susy}
\eea
It follows from (\ref{transgress}) that the 2-form field strength
is given by $\hat F_{mz} = G_m = \lambda\varepsilon_{mn}\, y^n/r^2$.
The functions $A$ and $B$ appearing in the $D=9$ solitonic 6-brane
metric, the Kaluza-Klein scalar $\varphi$, and the $D=10$ dilaton $\phi$
are given in terms of $\tilde\phi$ by $A=0$, $B=\ft12\tilde\phi$,
$\a\varphi=-\ft1{16}\tilde\phi$, $\phi=\ft34\tilde\phi$.  The Kaluza-Klein
vector potential ${\cal A}_{\sst M}$ is equal to zero.  With these, and
the expressions (\ref{spincon}) for the $D=10$ spin connection appearing in
$\hat D_{\hat\sst{M}}$ in terms of the $D=9$ spin connection and $\varphi$,
it is now straightforward to substitute the
oxidized solution into the fermionic transformation rules given in eqn
(\ref{d10susy}).  We find that half the supersymmetry is preserved if
$\varepsilon$ satisfies the conditions
\beq
\varepsilon = e^{-\ft1{32}\tilde\phi}\, \varepsilon_0\ ,\qquad
\hat\Gamma_{{\underline m}{\underline n}}\, \varepsilon_0 =
- \epsilon_{mn}\, \hat\Gamma_{\underline z}\, \hat\Gamma^{11}\,
\varepsilon_0\ ,
\eeq
where $\varepsilon_0$ is a constant spinor. Having demonstrated in $D=10$
that the non-isotropic oxidation (\ref{ox6brane}) of the 6-brane preserves
half of the type IIA, $D=10$ supersymmetry, it follows that the 6-brane
solution itself in $D=9$ also preserves half of the $N=2$, $D=9$
supersymmetry.

\section{$D\le 8$ supergravity}

     As one descends through the dimensions, starting at $D=11$, one
encounters various stainless brane solutions.  First of all, they occur if
the supergravity theory in a given dimension cannot be obtained from
dimensional reduction.   This happens in $D=11$, and $D=10$ for type IIB
supergravity.  A second reason for the occurrence of stainless brane solutions
is if no supergravity theory in the next higher dimension has the necessary
antisymmetric tensor field strength.  The above two reasons account for all the
stainless brane solutions in $D=11$ and $D=10$,  and the stainless solitonic
6-brane in $D=9$.  By the time one has reached $D=9$, all possible degrees
$n\le D/2$ for antisymmetric tensor field strengths have occurred. Because of
this,  any further stainless brane solutions in $D\le 8$ will arise only for
the third of the reasons we discussed in section 2.2, namely, that the $\Delta$
values for the exponential dilaton prefactors of the relevant antisymmetric
tensors in the higher and lower dimensions differ.   This phenomenon already
occurred for the 2-form field strength in $D=9$, giving rise to the stainless
elementary particle and solitonic 5-brane, as we discussed in the previous
section.

      In view of the above considerations, it is not surprising that
further stainless brane solutions in $D\le 8$ are relatively sparse.
However, we shall not attempt in this paper to give a full classification
of the super $p$-brane solutions for $D\le 8$. In $D=8$ and $D=7$, there
are stainless elementary particle solutions.  These solutions arise using
the 2-form field strength with $\Delta = 2$.  They are stainless since all
the 3-form field strengths in one dimension higher have $\Delta = 4$.

     In $D=6$, analogously to the cases of $D=8$ and $D=7$, there is also a
stainless elementary particle obtained from the 2-form field strength with
$\Delta =2$, which is part of the supergravity multiplet in $N=2$, $D=6$
supergravity.  In $N=1$, $D=6$ supergravity, on the other hand, there
exists a self-dual 3-form field strength, and there is no dilaton.  As we
discussed in section 2.2, brane solutions are still given by eqn
(\ref{solution1}), with $a$ set to zero, even in the absence of the
dilaton.   Thus this self-dual string solution is equivalent to the case
where $\Delta = 2$, with the metric given by (\ref{metrsol}).   Since there
is no supergravity theory in $D=7$ that contains a 3-form or 4-form field
strength with $\Delta=2$, the self-dual string in $D=6$ is stainless.

    The existence of a 3-form with $\Delta=2$ in $D=6$ implies that there is
no further stainless elementary particle in $D\le 5$ that arises from the
2-form field strength with $\Delta = 2$.   However, in $N=1$, $D=5$
supergravity, a new value of $\Delta$ for the 2-form field strength arises,
namely $\Delta = 4/3$.  This reflects the fact that there is no dilaton in
the theory. This 2-form field strength accordingly gives rise to a
stainless elementary particle and a stainless solitonic string \cite{GHT},
with metrics given by eqn (\ref{metrsol}).  To see how this works, we can
carry out the Kaluza-Klein dimensional reduction of $N=1$, $D=6$
supergravity.  Its bosonic sector comprises just the metric tensor and the
self-dual 3-form field strength mentioned above.  Since there is no
covariant action for this theory, we must instead implement the dimensional
reduction on the equations of motion themselves.  The bosonic equations of
motion are given by
\beq
\hat R_{\hat\sst{M} \hat\sst{N}} = \ft14 \hat F_{\hat\sst{M}\hat\sst{P}
\hat\sst{Q}}\hat F_{\hat\sst{N}}{}^{\hat\sst{P}\hat\sst{Q}}\ ,\qquad
{\hat F}_{\hat\sst{M}\hat\sst{N}\hat\sst{P}}= *{\hat
F}_{\hat\sst{M}\hat\sst{N}\hat\sst{P}} \ .\label{d6eqmo}
\eeq
The Kaluza-Klein ansatz for the metric and the antisymmetric tensor are
given by (\ref{kkans}) and (\ref{transgress}) as usual, but now, the
self-duality condition $\hat F= *\hat F$ implies that the lower-dimensional
2-form and 3-form field strengths $G_2$ and ${G'}_3$ are related:
\beq
{G'}_{\sst{ABC}}= \ft12 e^{4\a\varphi}\,
\epsilon_{\sst{ABCDE}}\, G^{\sst{DE}}\ ,
\eeq
where $\a$, given by (\ref{alphaval}), takes the value $\a=1/(2\sqrt6)$.
Substituting these ans\"atze into the 6-dimensional equations of motion
(\ref{d6eqmo}), and making use of the expressions (\ref{ricten}) for the
Ricci-tensor components, we obtain the 5-dimensional equations of motion
\bea
R_{\sst{AB}} &=& \ft12 \del_{\sst{A}}\varphi\, \del_{\sst{B}}\varphi +
\ft12 e^{-8\a\varphi}\,
({\cal F}^2_{\sst{AB}} -\ft16 {\cal F}^2 \eta_{\sst{AB}}) +
e^{4\a\varphi}\, (G^2_{\sst{AB}}
-\ft16 G^2 \eta_{\sst{AB}})\ ,\nonumber\\
\nabla^{\sst{B}}\Big( e^{-8\a\varphi}\, {\cal F}_{\sst{AB}} \Big) &=&
\ft14 e^{3\a\varphi}
\epsilon_{\sst{ABCDE}} G^{\sst{BC}} G^{\sst{DE}}\ ,\label{d5eqmo}\\
 \square \varphi &=& 2\a e^{4\a\varphi}\, G^2 - 2\a e^{-8\a \varphi}\, {\cal
F}^2\ .\nonumber
\eea

     We see that we may consistently truncate these fields to those of
minimal $D=5$ supergravity, whose bosonic sector comprises just the metric
and a 2-form field stength, by setting $\varphi=0$ and ${\cal F}_{\sst{AB}}
=G_{\sst{AB}}$. Defining $F_{\sst{AB}}\equiv \sqrt3{\cal F}_{\sst{AB}}=
\sqrt3 G_{\sst{AB}}$, we find that the
equations of motion for the remaining fields can be derived from the
Lagrangian
\beq
{\cal L} =e R -\ft14 e F^2 -\ft1{12\sqrt3}\epsilon^{\sst{MNPQR}}\,
F_{\sst{MN}} F_{\sst{PQ}} A_{{\sst R}}\ ,\label{d5lag}
\eeq
where $F_{\sst{MN}}=2\del_{[\sst{M}} A_{\sst{N}]}$.  This Lagrangian
describes the bosonic sector of minimal $D=5$ supergravity. We see
that a 2-form field strength with a new value of $\Delta$ has emerged in
the descent to five dimensions, namely $\Delta=\ft43$ and hence $a=0$.  It
follows that brane solutions in minimal $D=5$ supergravity, which make use
of this 2-form field strength, cannot be oxidized to give isotropic brane
solutions in any higher dimension.  In this way, we obtain the stainless
elementary particle and solitonic string solutions referred to above.
Their metrics are given by (\ref{metrsol}), with $d=1$, $\dt=2$ and $d=2$,
$\dt=1$ respectively, where $\Delta=\ft43$.

     To check the unbroken supersymmetry of these solutions, we need the
gravitino transformation rule in $D=5$ simple supergravity, which reads
\beq
\delta \psi_{\sst M} = D_{\sst M}\varepsilon -\ft{\im}{8\sqrt3}
F_{\sst{NP}}\Big(\Gamma_{\sst
M}{}^{\sst{NP}} -4 \delta_{\sst M}^{\sst N} \Gamma^{\sst P} \Big)\varepsilon
\ .\label{d5susy}
\eeq
For the solitonic string, we decompose the $D=5$ gamma matrices in the
$2+3$ split $\Gamma^\mu=\gamma^\mu\otimes \oneone$, $\Gamma^m=\gamma_3
\otimes \gamma^m$, where $\gamma_0 \gamma_1 =\gamma_3$ on the brane volume,
and $\gamma_1\gamma_2\gamma_3=\im$ in the transverse space.  Thus we have
\bea
\delta\psi_\mu &=& \fft12 \del_m A\, e^{A-B}\, \gamma_\mu\gamma_3\otimes
\gamma_m \varepsilon +\fft{\lambda}{4\sqrt3}\, e^{A-2B}\, \fft{y^m}{r^3}\,
\gamma_\mu\otimes \gamma_m \varepsilon\ ,\nonumber\\
\delta\psi_m &=& \del_m\varepsilon +\fft{\im}{2}\del_n B\, \epsilon_{mnp}
\oneone\otimes\gamma_p \varepsilon + \fft{\lambda}{4\sqrt3} e^{-B}\,
\fft{y^m}{r^3}\, \gamma_3\otimes\oneone\, \varepsilon -\fft{\im
\lambda}{2\sqrt3}e^{-B}\, \fft{y^n}{r^3}\, \gamma_3\otimes\gamma_p\,
\varepsilon \ .\label{d5solsusy}
\eea
Substituting the solitonic string solution, which, from (\ref{abphirel})
in the limit $a=0$ has $A'=-\ft12 B'= -\lambda/(2\sqrt3) e^{-B}r^{-2}$, we
find that half of the supersymmetry is preserved provided that
\beq
\varepsilon =e^{\ft12 A}\, \varepsilon_0\ , \qquad \gamma_3\otimes\oneone
\, \varepsilon_0=\varepsilon_0\ ,\label{stringeps}
\eeq
where $\varepsilon_0$ is a constant spinor.

     For the elementary particle, we decompose the $D=5$ gamma matrices in
the $1+4$ split $\Gamma^0=\im \gamma_5$, $\Gamma^m= \gamma^m$, where
$\gamma_5=\gamma_1\gamma_2\gamma_3\gamma_4$.  The supersymmetry
transformation rule (\ref{d5susy}) becomes
\bea
\delta\psi_0 &=& \ft{\im}{2} e^{A-B}\,\del_m A\, \gamma_m\gamma_5\varepsilon
-\ft{\im}{2\sqrt3} e^{C-B}\, \del_m C\, \gamma_m \varepsilon\ ,\nonumber\\
\delta\psi_m &=& \del_m\varepsilon +\ft12 \del_n B\, \gamma_{mn}\varepsilon
+\ft{1}{4\sqrt3} e^{C-A}\, \del_n C\, \gamma_{mn}\gamma_5\varepsilon
-\ft{1}{2\sqrt3} e^{C-A}\, \del_m C\, \gamma_5\varepsilon\ .
\label{d5susyele}
\eea
Substituting the elementary particle solution, which, from
(\ref{abphirel}) in the limit $a=0$ has $A'=-2B'= (1/\sqrt3) e^{C-A}\, C'$,
we find that half of the supersymmetry is preserved provided that
\beq
\varepsilon=e^{\ft12 A}\, \varepsilon\ ,\qquad \gamma_5\, \varepsilon_0=
\varepsilon_0\ ,
\eeq
where $\varepsilon_0$ is a constant spinor.

It is interesting to note that there
are in total {\it three} inequivalent solitonic string solutions in $D=5$,
namely the stainless example we have just derived, a rusty string that
oxidizes to our stainless 5-brane in $D=9$, and another rusty string that
oxidizes to the stainless 6-brane in $D=10$.  Their metrics in $D=5$ are
given by (\ref{metrsol}) with $d=2$ and $\dt=1$, by taking $\Delta=\ft43$,
$\Delta=2$ and $\Delta=4$ respectively.  Upon dimensional reduction to
$D=4$, they give rise to particles with $a=1/\sqrt3$, 1 and
$\sqrt 3$ respectively.  These correspond to the black hole solutions of
$D=4$ string theory (see, for example, \cite{dkl}).\footnote{We are
grateful to J. Rahmfeld for drawing our attention to the black hole
solutions in the $D=4$ string.}

\section{Zero modes}

     In the previous sections, we described stainless $p$-brane solutions in
various dimensions.  The complete set of brane solutions is thus given
by those solutions together with their descendants {\it via} Kaluza-Klein
double dimensional reduction.   All of these solutions break half of the
supersymmetry.

     Each broken supersymmetry transformation in a $p$-brane solution gives
rise to a corresponding fermionic Goldstone zero mode.  There will also be
bosonic zero modes associated with the breaking of local bosonic gauge
symmetries by the non-vanishing $p$-brane background solution.  These will
certainly include the translational zero modes corresponding to the broken
constant general coordinate transformations $\delta y^m=c^m$ in the
space transverse to the $p$-brane world volume.  Thus there will be
$D-d=\dt+2$ such bosonic zero modes.  Since supersymmetry remains partially
unbroken by the solution, it follows that the fermionic and bosonic zero
modes must form supermultiplets under the remaining unbroken supersymmetry.
 In particular, there must be equal numbers of fermionic and bosonic
zero-mode degrees of freedom.

     The matching of the zero modes for the bosonic and fermionic
fields is straightforward in the case of the elementary membrane in $D=11$
and the solitonic 5-brane in $D=10$, and also for all their descendants {\it
via} dimensional reduction. In all of these cases, the number of translational
zero modes is precisely  the same as that of the fermionic zero
modes, {\it i.e.}\ the number of on-shell fermionic zero-mode degrees of
freedom.  Thus for the supermembrane in $D=11$, there are $8=32/2/2$
fermionic zero modes, where the original 32 components of the supersymmetry
parameter in $D=11$ are halved once to arrive at the number of on-shell
degrees of freedom, and halved again because half of the supersymmetries
are broken.  The membrane solution breaks translational invariance in the
$y^m$ directions, giving rise to $8=11-3$ bosonic zero modes.  The same
counting of $8+8$ degrees of freedom holds for the dimensional reduction to
the string in $D=10$.  For the solitonic 5-brane in $N=1$, $D=10$
supergravity, there are $4=16/ 2/2$ fermionic zero modes, and $4=10-6$
bosonic translational zero modes.  This matching of $4+4$ degrees of
freedom holds for the various stages of dimensional reduction all the way
down to the string in $D=6$ and the superparticle in $D=5$.

     In all the other brane solutions, the number of translational zero modes
is less than the number of fermionic zero modes.  Since we know that the
remaining unbroken symmetry guarantees a matching of the bose and fermi zero
modes, it follows that the there must be further bosonic zero modes
associated with these solutions.  They arise from the breaking of
antisymmetric tensor gauge symmetries.  The simplest way to find these
additional bosonic zero modes is first to construct the fermionic zero
modes, and then to obtain their bosonic partners by transforming them under
the remaining unbroken supersymmetries.

     We shall carry out this procedure first for the stainless solitonic
5-brane in $N=1$, $D=9$ supergravity, which we constructed in section 4.1.
This solution has four fermionic zero modes; however, it has only $9-6=3$
translational zero modes.  As we shall see, there is one further bosonic
zero mode associated with the breaking of the gauge invariance of the 2-form
field strength that takes a non-zero value in the background solution.

     The fermionic supersymmetry transformations in the background of the
solitonic 5-brane are given by eqn (\ref{d9soltr}).  As we discussed in
section 4.1, these variations vanish for spinors $\varepsilon$ satisfying
(\ref{d9epso}), which includes a chirality condition.  They correspond to
the unbroken supersymmetry generators. The broken generators, on the other
hand, correspond to supersymmetry parameters $\eta$ that have the opposite
chirality under the $\gamma_7$ matrix on the world volume. Specifically, we
shall consider spinors $\eta$ given by
\beq
\eta= e^{-\ft12 A}\, \eta_0\ ,\qquad \gamma_7\otimes\oneone\, \eta_0 =-\eta_0\
,
\label{etasol}
\eeq
where $\eta_0$ is constant.  This choice is motivated by the
simplifications to the fermionic zero-mode structure that result.  Note
that any other asymptotically constant spinors with the same $\gamma_7$
eigenvalue could equally well have been chosen.  These would lead to
zero-modes differing from ours by pure gauge transformations whose
parameters die off at infinity.  With our choice, it follows from
(\ref{d9soltr}) that the purely bosonic 5-brane soliton background varies
into the following fermionic configuration:
\bea
\chi&=& \ft{1}{\sqrt2} e^{-\ft12 A-B} \, \del_m\phi\,
\oneone\otimes\gamma_m\, \eta_0\ ,\nonumber\\
\psi_\mu&=& -\del_m A\, e^{\ft12 A -B}\,\gamma_\mu\otimes\gamma_m\, \eta_0 \ ,
\label{d95brfermzm}\\
\psi_m&=& \del_n B\, e^{-\ft12 A}\, \oneone\otimes\gamma_{mn}
\eta_0\ .\nonumber
\eea
These, then, describe the four fermionic zero modes, parametrised by the
eight independent spinors $\eta_0$ that satisfy the chirality condition
given in (\ref{etasol}).  (Recall that for $d\ge3$, the count of on-shell
fermionic degrees of freedom is half that of the off-shell spinor fields.)
We can substitute these spinors into the bosonic transformation rules
(\ref{d9bostr}), taking the supersymmetry parameter $\varepsilon$ to be one
of the eight unbroken generators given by (\ref{d9epso}), in order to
obtain the bosonic superpartners of the fermionic zero modes.

     Carrying out this procedure, we find the following non-vanishing
results for the bosonic zero modes:
\bea
\delta\phi &=& \delta_{\rm Diff}\, \phi\, \nonumber\\
\delta e^{\underline\mu}{}_\nu &=& \delta_{\rm Diff}\,
e^{\underline\mu}{}_\nu + \Omega^{\underline\mu}{}_{\underline\rho} \,
e^{\underline \rho}{}_\nu\ ,\nonumber\\
\delta e^{\underline m}{}_n &=& \delta_{\rm Diff}\, e^{\underline m}{}_n +
\Omega^{\underline m}{}_{\underline p} \,
e^{\underline p}{}_n\ ,\label{zmodes}\\
\delta A_m &=& \delta_{\rm Diff}\, A_m + \del_m\widetilde\Lambda\ ,
\nonumber\\
\delta B_{mn} &=& \delta_{\rm Diff}\, B_{mn} -\ft12 \widetilde\Lambda F_{mn}
+ \del_{[m} \widetilde\Lambda_{n]}\ ,\nonumber
\eea
where $\delta_{\rm Diff}$ denotes a diffeomorphism transformation,
$\delta_{\rm Diff}\, V_m = c^n\del_n V_m + \del_m c^n\, V_n$, {\it etc}., and
the composite transformation parameters are given by\footnote{These results
could be obtained from the commutator algebra of the local supersymmetry
transformations. To see this, note that the bosonic zero modes $b$ can be
written in terms of the femionic zero modes $f$ as $b=\delta_\epsilon f$.
However, $f$ can be written as $f=\delta_\eta B$, where $B$ represents the
supersymmetric bosonic background fields. Since $\delta_\epsilon B=0$, we can
write $b=[\delta_\epsilon,\delta_\eta] B= \delta_c B+\delta_{\Lambda_{0}}
B+\delta_{\Lambda_1} B+\delta_\Omega$, where $c$, $\Lambda_{0}$, $\Lambda_{1}$
and $\Omega$ are the composite parameters for the general coordinate, abelian
gauge, antisymmetric gauge and Lorentz transformations, respectively.}
\bea
c^m &\equiv& \im e^{-B} {\bar \varepsilon}_0 \oneone\otimes \gamma_m\eta_0\ ,
\nonumber\\
\Lambda &\equiv& \sqrt2 e^{-A}\, {\bar \varepsilon}_0\eta_0\ ,\nonumber\\
\Lambda_m &\equiv& -\ft74 e^{-2A}\, c_m\ ,\label{zmodeparms}\\
\widetilde\Lambda &\equiv& \Lambda -c^m A_m\ , \qquad \widetilde \Lambda_m
\equiv \Lambda_m - \Lambda A_m + 2 c^p B_{mp}\ \nonumber\\
\Omega^{\underline \mu}{}_{\underline\rho} &\equiv& \im \bar\varepsilon_0
\gamma^{\mu}{}_\rho\otimes \gamma_m \eta_0 \,e^{-B}\, \del_m A\ ,\nonumber\\
\Omega^{\underline m}{}_{\underline p} &\equiv& \del_m c^p - \del_p c^m
-\epsilon^m{}_{pq}\, \bar\varepsilon_0 \eta_0 \,e^{-B}\,\del_q B\ .
\nonumber
\eea
The quantities $\widetilde\Lambda$ and $\widetilde\Lambda_m$ are gauge
transformation parameters for the potentials $A_m$ and $B_{mn}$ respectively;
$\widetilde\Lambda$ appears also in the $B_{mn}$ transformation because its
field strength is given by $G=dB + \ft12 A\wedge F$.  As one chooses different
constant spinors $\varepsilon_0$ and $\eta_0$, satsifying their respective
$\pm1$ chirality conditions under $\gamma_7$, the diffeomorphism parameter
$c^m$ and gauge transformation parameter $\Lambda_m$ jointly fill out a
3-dimensional space of gauge transformations that are constant and
non-vanishing as $r$ tends to infinity.  Likewise, $\Lambda$ describes a
further independent asymptotically constant non-vanishing gauge transformation.
 Taken together, we have the four independent bosonic zero modes of the
solitonic 5-brane in $D=9$ dimensions.  Note that $\Omega^{\underline
\mu}{}_{\underline \rho}$ and $\Omega^{\underline m}{}_{\underline p}$ are
parameters of Lorentz transformations that die off at infinity and thus do not
contribute to the true zero modes, which correspond to the broken generators of
the global asymptotic symmetry group.

     The zero-modes of the 5-brane in $D=9$ are thus properly balanced
between the bose and fermi sectors. With respect to the unbroken $N=1$, $D=6$
supersymmetry, they form a hypermultiplet, in the form where the four scalars
occur as an $SU(2)$ triplet plus a singlet. The spinor zero-modes form an
$SU(2)$-Majorana doublet. Note that while the zero-modes form a
supermultiplet under the unbroken supersymmetry in $d=6$, as they must, not
all of the scalars correspond to translational zero-modes. In the reduction
from 9 to 6 dimensions, only three translational modes occur, leaving one
more to arise from a different broken gauge symmetry. In the present case, this
extra scalar mode arises from the broken gauge symmetry of the $A_{\sst M}$
potential, {\it i.e.}\ the $\Lambda$ zero mode in
(\ref{zmodeparms}).\footnote{Note also that giving an expectation value to an
Abelian field strength can cause symmetry breaking in the present context,
unlike in an ordinary Yang-Mills context, because of the structure of the
linked gauge symmetry involving $A_{\sst M}$ and $B_{\sst{MN}}$.}

     The other $p$-brane solutions discussed in this paper all leave half
the original $D$-dimensional supersymmetry unbroken, and form appropriate
supermultiplets of the unbroken supersymmetry. As another example,
consider the string solution in $N=1$, $D=5$ supergravity that we discussed
in section 5. Of the original 8 real components of the supersymmetry
transformation, 4 are unbroken by the solution and 4 are broken, giving rise
to 4 fermionic zero-mode {\it fields}. Of the bosonic zero-modes, there are
obviously 3 corresponding to the broken translations. One more
scalar zero-mode arises from the broken gauge symmetry of the 2-form field
strength. In order to organise these into a supermultiplet of the unbroken
$d=2$ supersymmetry, one needs to recall one of the characteristic
features of $d=2$ supersymmetry. As one may see from eqn (\ref{stringeps}),
the surviving $d=2$ supersymmetry is {\it holomorphic}; it is in fact a
(4,0) supersymmetry. This supersymmetry relates the 4 fermionic
zero-mode fields to 4 holomorphic bosonic modes. The usual style of
counting zero-modes in $d\ge3$, in which the count of fermionic zero-modes
is taken to be half of the number of fermionic fields, is not particularly
convenient in $d=2$. In $d=2$, the bosons also need to be split into
holomorphic and antiholomorphic parts. Although our solitonic string
solution clearly will have both holomorphic and antiholomorphic components
of the bosonic zero-modes propagating according to the worldsheet equations
of motion, only one of these sectors becomes paired with the fermionic zero
modes in the (4,0) supermultiplet. The other sector remains unpaired as a
set of supersymmetric singlets.

\section{Discussion}

     In this paper, we have searched for supersymmetric $p$-brane solutions
of supergravity theories in diverse dimensions.  As these solutions occur
in families related by dimensional reduction, we have concentrated on the
maximal, or stainless, solution in each family.  In addition to the
previously-known examples, we have found a number of new solutions in $5\le
D\le 9$ dimensions.  (The lower dimensional bound arises here because we
have restricted our attention to solutions of supergravity theories, and
have not considered $p$-branes in theories with rigid supersymmetry.)  Our
new stainless solutions cannot oxidize isotropically to brane solutions in
higher dimensions.  Put another way, this means that these new solutions
cannot simply be viewed as dimensional reductions of previously-known brane
solutions in $D=11$ or $D=10$ dimensions.

     A question that we have not addressed so far concerns the world brane
actions that should describe the zero-mode fluctuations around the static,
isotropic solutions considered here.  From supersymmetry, one has detailed
knowledge of the supermultiplet structure of the zero modes that would
appear in a gauge-fixed world brane action.  In general, one expects that
such gauge-fixed actions should be extendable to spacetime supersymmetric
and Lorentz invariant actions by adding the appropriate additional
unphysical degrees of freedom and their associated local world volume gauge
symmetries.  For the four classic sequences of super $p$-branes, these
actions are generalisations of the $D=10$ superstring \cite{gsch,pol} or $D=11$
supermembrane \cite{berg} actions.  Very little is known about the
structure of the covariant world volume actions for any of the other
$p$-brane cases; this subject remains an important open problem.

     In the absence of detailed knowledge of the world volume action, some
information can be extracted if one assumes that the bosonic sector of the
action takes the general form of a Nambu-Goto action coupled to the
spacetime metric, dilaton, and a $d$-form gauge potential, with the
non-polynomiality removed by the introduction of a world-volume metric
$\gamma_{ij}$ as an auxiliary field \cite{bhd}:
\bea
I_{\rm brane} &=& \int d^d\xi \Big(-\ft12 \sqrt{-\gamma} \gamma^{ij}\,
\del_i X^{\sst M} \del_j X^{\sst N}\, g_{\sst{MN}}\, e^{b\phi}
+\fft{d-2}{2} \sqrt{-\gamma} \nonumber\\
&&- \fft{1}{d!}\, \epsilon^{i_1\cdots i_d}\,
\del_{i_1} X^{{\sst M}_1}\cdots \del_{i_d} X^{{\sst M}_d}\, A_{{\sst
M}_1\cdots {\sst M}_d}\Big) \ .\label{bract}
\eea
The exponent $b$ of the dilaton coupling in the first term is {\it a
priori} unknown.  One way of selecting a value for it in a number of cases
is by resorting to an argument based on a scaling symmetry of the pure
supergravity theory \cite{dkl}.  Under the constant rescalings
\beq
e^\phi\longrightarrow \lambda^\a\, e^\phi\ ,\qquad
g_{\sst{MN}}\longrightarrow \lambda^{2\beta}\, g_{\sst{MN}}\ ,\qquad
A_{\sst{M}_1\cdots \sst{M}_d}\longrightarrow \lambda^\gamma\,
A_{\sst{M}_1\cdots \sst{M}_d}\ ,\label{scale}
\eeq
the action given by (\ref{boslag}) scales by an overall factor
$\lambda^{\beta(D-2)}$, provided that we choose $\gamma=\beta d + a \a$.
Requiring that $I_{\rm brane}$ scale in the same way, one finds that the
parameters must be related by
\beq
\a=\fft{d\dt}{a(D-2)}\ ,\qquad \beta=\fft{d}{D-2}\, \qquad \gamma=d\ ,\qquad
b=\fft{a}{d}\ .\label{solpar}
\eeq
Substituting the ans\"atze (\ref{metrform}) and (\ref{eleans}) into
(\ref{bract}), one finds that the branewave equation following from
(\ref{bract}) implies \cite{dkl}
\beq
(e^C)'= (e^{dA + a\phi/2})'\ .\label{branewave}
\eeq
Comparing with the solution given by (\ref{solution1}) and (\ref{csol}),
one finds that $a$ must satisfy
\beq
a^2=4 -\fft{2d\dt}{D-2}\ .\label{dudelta}
\eeq
Thus requiring that the elementary brane solution should also satisfy the
branewave equation, with the parameter $b$ in (\ref{bract}) determined by
requiring the above scaling symmetry, the value of $a$ appearing in
(\ref{boslag}) is in all cases given by (\ref{avalue}) with $\Delta=4$.

     All of the antisymmetric tensors in $D=11$ and $D=10$ supergravities have
$\Delta=4$, as we have seen.  Thus the elementary $p$-brane solutions
associated to these antisymmetric tensors accord with the above discussion.
However, as we have seen, there are other values of $\Delta$ that also
occur in lower dimensional supergravity theories.  Elementary $p$-brane
solutions in such cases cannot have zero modes that are described by the
action (\ref{bract}) with the choice of parameter $b$ dictated by the
scaling symmetry.  An example is provided by the self-dual string in $D=6$,
for which the self-dual 3-form has $\Delta=2$.
In this case, since there is no action for $N=1$, $D=6$ supergravity, one
has to implement the scaling argument at the level of the equations of
motion.  The supergravity equations of motion themselves (\ref{d6eqmo}) are
invariant under the scaling $g_{\sst{MN}}\longrightarrow \lambda^{2\beta}
g_{\sst{MN}}$ and $G_{\sst{MNP}}\longrightarrow \lambda^{2\beta}
G_{\sst{MNP}}$, for arbitrary $\beta$.  However, the energy-momentum tensor
$T_{\sst{MN}}(x)= -\int d^2\xi \sqrt{-\gamma}\gamma^{ij} \del_i X^{\sst P}
\del_j X^{\sst Q}\, g_{\sst{MP}} g_{\sst{NQ}}\delta^6(x-X)/\sqrt{-g}$
scales with a factor $\lambda^{-2\beta}$.  Thus the coupled
supergravity-string equations break the scaling symmetry.  In fact all of
the new stainless elementary $p$-branes, which have $\Delta=2$ or
$\Delta=\ft43$, exhibit a similar breaking of the scaling symmetry in their
couplings.

     The ultimate significance of the scaling symmetry used in the above
arguments remains unclear to us. Are elementary $p$-brane solutions that
respect the scaling symmetry in their couplings more fundamental than
others? We do not have an answer to this question at present. We would
remark, however, that couplings that break scaling symmetries present in
pure uncoupled gravity and supergravity theories are not at all uncommon.
Take, for example, a massless charged particle coupled to Einstein-Maxwell
theory in $D=4$. The Einstein-Maxwell action $I=\int d^4x (e R -\ft14 e
F^2)$ in the absence of the particle coupling scales as $I\longrightarrow
\rho^2 I$ under $g_{\mu\nu}\longrightarrow\rho^2 g_{\mu\nu}$,
$A_\mu\longrightarrow \rho A_\mu$. But once the particle is coupled in,
{\it via} the standard worldline-reparametrisation invariant action $\int
d\tau (e^{-1}\dot X^\mu\dot X^\nu g_{\mu\nu} + A_\mu\dot X^\mu)$, the
scaling symmetry is broken by the electromagnetic coupling $A_\mu\dot
X^\mu$.

     Another reason why the status of the scaling symmetries remains in
doubt comes from quantisation. The supergravity theories arising as
low-energy effective field theories of superstring theories have field
equations determined {\it via} the beta functions from the requirement that
the string's conformal invariance be preserved. The leading terms of these
effective field equations reproduce standard supergravity field equations,
but there are also an infinite series of quantum corrections, all of which
break the scaling symmetries.

      Nonetheless, it is intriguing that a purely bosonic discussion
based upon the coupling of Nambu-Goto-type actions and preservation of
scaling symmetries fixes dilaton couplings to antisymmetric tensor
gauge fields in a way that agrees with many of the couplings actually found in
supergravity theories ({\it i.e.}\ the $\Delta=4$ couplings).

      In concluding, we shall summarise the results that we have obtained
in this paper in a revised brane-scan, in which we plot only the stainless
members of each $p$-brane family.  In accordance with our discussion at the
end of section 2.1, we may, without loss of generality, consider only the
versions of the various supergravity theories where all the antisymmetric
tensors have degrees $n\le D/2$, since no further inequivalent elementary
or solitonic brane solutions arise from dualised versions of the theories.
Accordingly, in the stainless brane-scan, we denote solutions of the $n\le
D/2$ theories that are elementary by open circles, solutions that are
solitonic by solid circles, and self-dual solutions by cross-hatched
circles.  The dashed lines extending diagonally downwards from the various
points on the brane scan indicate that each stainless solution gives rise
to its own set of dimensionally-reduced descendants.

\newpage
\begin{center}
{\bf \Large Brane Scan of Stainless Supergravity Solutions}
\vskip1cm
\epsfbox{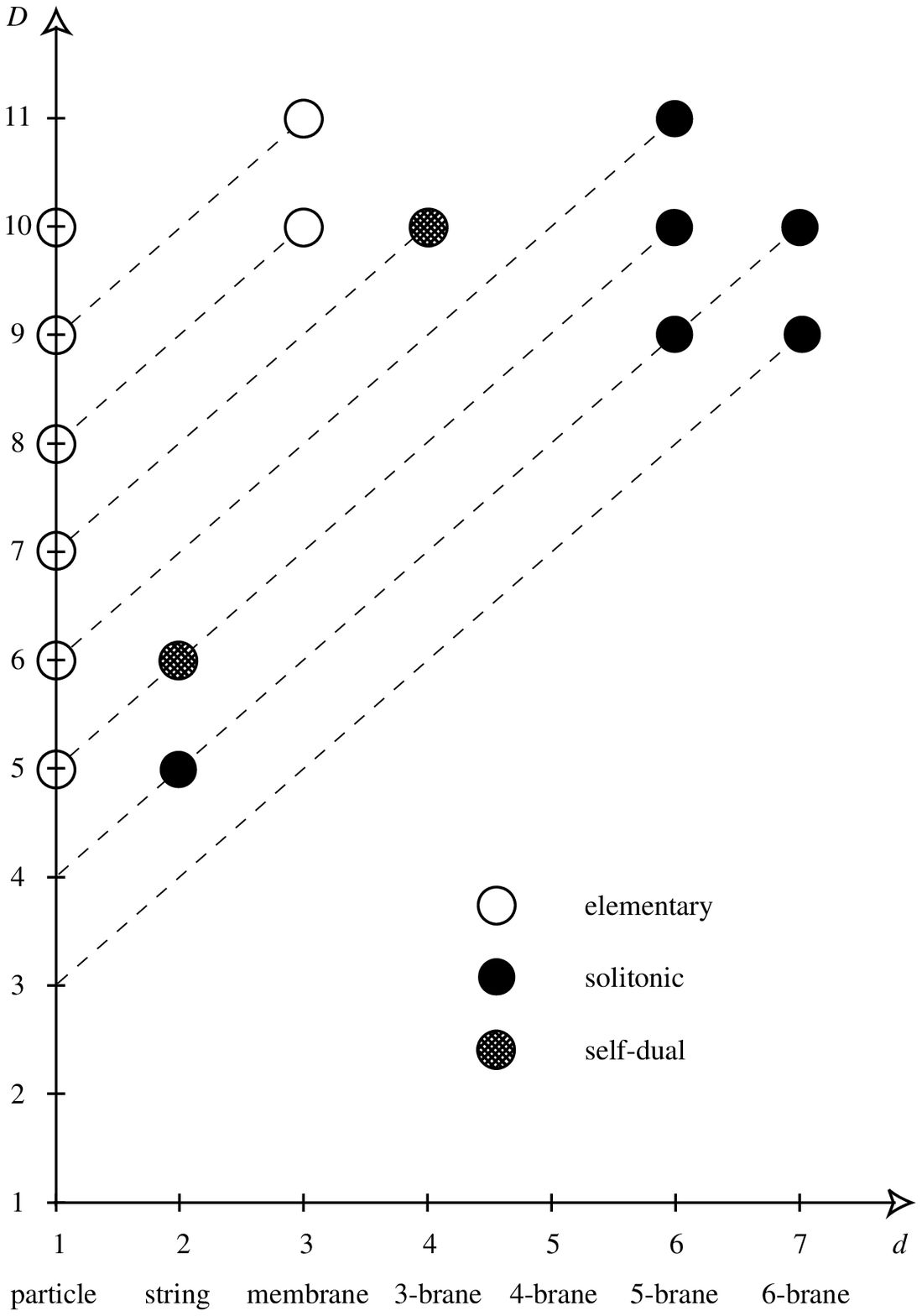}
\bigskip\bigskip
(For supergravity theories in their $n\le D/2$ versions)
\end{center}

\section*{Acknowledgements}

     H.L., C.N.P. and K.S.S. thank SISSA, Trieste, and E.S. thanks ICTP,
Trieste, for hospitality during the course of this work.

\end{document}